%% file: manuscript.tex
\documentclass{article}

\usepackage{arxiv}

\usepackage[utf8]{inputenc} 
\usepackage[T1]{fontenc}    
\usepackage{hyperref}       
\usepackage{url}            
\usepackage{booktabs}       
\usepackage{amsfonts}       
\usepackage{natbib}
\usepackage{nicefrac}       
\usepackage{microtype}      
\usepackage{lipsum}
\usepackage{graphicx}
\usepackage{listings}
\usepackage{placeins}
\usepackage{longtable}
\usepackage{booktabs}
\usepackage{authblk}

\graphicspath{ {./images/} }

\title{Structured AI Demonstrations and Student LLM Use in Engineering Mechanics: Study Design and Preliminary Results }

\author[1]{Shuang Geng}
\author[2]{Helen Lallos-Harrell}
\author[3]{Jiya Ashar}
\author[4]{Thomas J. McKenna}
\author[1]{Annwesa Dasgupta}
\author[5]{Caleb Farny}
\author[5]{Emma Lejeune\thanks{Corresponding author: Emma Lejeune, \texttt{elejeune@bu.edu}} $\,$}

\affil[1]{Institute for Excellence in Teaching and Learning, Boston University}
\affil[2]{Department of Economics and Department of Political Science, Boston University}
\affil[3]{Bioinformatics Program, Boston University}
\affil[4]{Wheelock College of Education \& Human Development, Boston University}
\affil[5]{Department of Mechanical Engineering, Boston University}

\begin{document}
\maketitle
\begin{abstract}
The rapid integration of large language models (LLMs) into undergraduate education presents an urgent challenge for engineering instructors. Despite widespread student adoption, there remains a critical lack of domain-specific empirical evidence to guide pedagogical policies and classroom interventions. This manuscript presents a descriptive study design and preliminary findings from an undergraduate engineering mechanics course conducted in Spring 2026. We detail a reproducible survey instrument used to capture student AI usage patterns, attitudes, and verification practices, which are subsequently linked to academic performance metrics. Additionally, we document a deployable sequence of nine structured, instructor-led AI demonstrations designed to model strategic LLM delegation and evaluation. While our preliminary data highlight shifting student behaviors and complex relationships between AI reliance and course outcomes, the primary contribution of this work is the provision of an open-access methodological framework. By making our complete study design, survey tools, and demonstration materials publicly available, we urge other engineering educators to collect and share similar empirical data. Navigating this unprecedented technological shift will require a collaborative, evidence-based approach to fully understand its long-term impacts on student learning.
\end{abstract}


\section{Introduction}

Since the public release of ChatGPT in November 2022, large language models (LLMs) have rapidly become part of the undergraduate learning environment \citep{stephenson2026student}. Early public-facing systems had important limitations for engineering education. For example, they could not accept image uploads through the web interface, which constrained their usefulness for diagram-rich engineering problems, and they often produced incorrect or fabricated solutions even for problems that could be stated entirely in text \citep{qadir2023engineering}. By the time of the present study (January-May 2026), however, general-purpose AI tools had become substantially more capable \citep{minaee2024large,yue2024mmmu}, more accessible, and more deeply embedded in students’ academic workflows \citep{stephenson2026student,freeman2024provide,freeman2025student}. For instructors in engineering courses, this shift creates a practical and urgent curricular question: how, if at all, should these tools be integrated into engineering education? Plausible responses range from prohibiting or discouraging AI use, to treating LLMs as supplemental learning resources, to redesigning core curricula around AI-supported problem solving \citep{barba2025experience}. Despite strong opinions across this spectrum, there remains limited empirical evidence to guide concrete instructional decisions, particularly in domain-specific engineering problem-solving courses \citep{ariza2025generative}. The goal of this manuscript is to begin addressing that gap through a descriptive, course-specific study of undergraduate engineering mechanics students’ AI use, attitudes, verification practices, and course outcomes.

Early evidence on how LLM use affects learning is mixed, and the emerging pattern is that outcomes depend less on the technology than on how students use it. A meta-analysis of 69 experimental studies finds positive average effects of ChatGPT use on academic performance but no effect on self-efficacy, with outcomes varying by instructional integration \citep{deng2025does}. The randomized evidence points in both directions. In one trial, unrestricted access to a GPT-4 chat interface improved high school students' practice performance but significantly harmed their subsequent unassisted exam performance, while a tutor variant with pedagogical guardrails largely eliminated that harm \citep{bastani2025generative}. In another, a purpose-built AI tutor in a large introductory university physics course roughly doubled learning gains \citep{kestin2025ai}. That tutor was designed around the same principles as the active-learning instruction it replaced. Naturalistic classroom data show the same split within a single cohort. A semester-long log of 16,851 student-ChatGPT conversations in a university artificial intelligence course found that prompting the tool for conceptual understanding and coding help correlated with better performance on assignments and exams, while using it to write reports or circumvent learning objectives correlated with lower exam outcomes \citep{mcnichols2026studychat}.
Consistent with these patterns, a second meta-analysis, covering 35 experimental studies, reports a moderately positive average effect on learning outcomes (Hedges' $g = 0.67$) that varies significantly by instructional mode, again locating the effect in how the tool is used rather than in the tool itself \citep{wu2026chatgpt}.
The primary base is now large enough to sustain multiple independent syntheses: quantitative meta-analyses \citep{deng2025does, wu2026chatgpt}, broad systematic reviews of generative AI in higher education \citep{batista2024generative}, and discipline-specific reviews and studies in fields adjacent to engineering such as science education \citep{almasri2024exploring} and mathematics \citep{fardian2025integrating}. These converge on a field expanding quickly but still short on rigorous, domain-specific evidence, precisely the gap this study targets.

In the context of engineering problem-solving, the educational utility of large language models is inextricably linked to both their technical capabilities and student learning behaviors \citep{zhou2026engibench,wang2025physunibench}. LLMs now pass introductory physics coursework \citep{kortemeyer2023could} and perform at passing levels on Fundamentals of Engineering (FE) exams \citep{pursnani2023performance,frenkel2024chatgpt}.\footnote{Passing-level performance on the FE exam indexes a narrow, closed-form problem-solving format and should not be read as a general measure of engineering capability.} Notably, LLM capability remains weakest on visual representations which is highly relevant for engineering problems. In work that now predates current models, ChatGPT was shown to perform poorly on a test of interpreting kinematics graphs \citep{polverini2024performance}. Consistent with this, a highly related study in an introductory statics course using a GPT-4-era textbook chatbot found that most students did not use it to seek answers; verbatim problem-pasting was a minority behavior that clustered late at night, outside office hours, and students who did paste problems often found the answers unreliable and shifted toward using the tool for methodology and conceptual clarification \citep{wyszynski2025ace, defrancisis2025exploring, wyszynski2025wip}. These findings predate the current generation of vision-capable, reasoning-oriented models, leaving open how today's substantially more capable tools are used and verified in engineering coursework.
At present, the literature lacks naturalistic, within-semester evidence of how engineering students use and verify these tools in a core problem-solving course, how that use relates to performance, and whether light-touch instruction shifts their behavior.

This study is aimed at this gap and is focused on three research questions:
\begin{itemize}
\item How do undergraduate engineering students use LLMs in their coursework, and how do their use patterns, attitudes, and verification practices vary over the course of a semester?
\item How do students' LLM use patterns relate to course performance?
\item What associations, if any, exist between brief, recurring instructor-led AI demonstrations and students' LLM-related practices, attitudes, and course outcomes?
\end{itemize}
In addressing these questions, this work provides a domain-specific empirical snapshot of undergraduate engineering mechanics students' generative AI use, attitudes, and verification practices in Spring 2026, linked to course performance so that AI-related behaviors can be interpreted in relation to academic outcomes rather than as isolated attitudes. It also documents a reproducible study design, survey instrument, and set of structured AI demonstrations that can be adapted by other instructors. This study is the first semester of an ongoing AI-in-instruction effort. The present analysis is primarily descriptive: modest sample sizes and non-random section assignment preclude strong causal claims about the demonstrations, which here are reported as context rather than as a treatment effect under evaluation. We offer these results, study design, and survey instrument as transparently as possible in order to enable reproducibility and to serve as an empirical starting point for understanding how students are already incorporating LLMs into engineering problem solving.

\section{Methods}

This Section outlines the methodological framework of the study. We first describe the instructional setting, participant demographics, and institutional generative AI policies (Section \ref{sec:course_context}). Next, we detail the quasi-experimental pre- and post-semester study design (Section \ref{sec:design}) and acknowledge its inherent limitations (Section \ref{sec:limitations}). The core intervention, consisting of nine structured classroom AI demonstrations, is outlined in Section \ref{sec:intervention}, with complete demonstration workflows and the Socratic tutor system prompt provided in Appendices \ref{apx:demos} and \ref{apx:prompt}. Our data collection measures, which include academic performance metrics and the full survey instrument (Appendix \ref{apx:survey}), are defined in Section \ref{sec:measures}. Finally, we review our data preparation and de-identification procedures (Section \ref{sec:data_prep}), followed by our quantitative and qualitative analysis strategies (Section \ref{sec:analysis}), with the complete qualitative codebooks and the LLM-assisted machine coding protocol thoroughly documented in Appendices \ref{apx:codebook} and \ref{apx:llm_coding}.

\subsection{Course Context and Participants}
\label{sec:course_context}

This study was conducted in EK301 (Engineering Mechanics I) at Boston University during the Spring 2026 semester. EK301 is a required undergraduate engineering core course taken by all College of Engineering majors and Foundation Phase LEAP students, with PY211 (Introductory Physics) as a prerequisite and MA225 (Multivariate Calculus) and EK125 (Introduction to Programming for Engineers) as corequisites. The course covers the statics of particles and rigid bodies in two and three dimensions, internal forces and moments, analysis of trusses (via the methods of joints and sections), frames, dry friction, distributed forces, centroids, shear and bending moment diagrams, and virtual work, using Hibbeler's \textit{Engineering Mechanics: Statics} as the required textbook. In Spring 2026, EK301 was offered across three lecture sections taught by three different instructors, with 196 students enrolled across all sections. All sections shared a common syllabus, weekly homework assignments, and a semester-long truss design project in which teams of three students characterized acrylic bar buckling behavior in an early-semester lab, then designed, analyzed, built, and load-tested a truss subject to design constraints. Weekly quizzes and exams (two midterms and a cumulative final) varied problems across sections to prevent cheating but followed a common structure and weighting (each exam was worth 20$\%$ of the final grade). One section, hereafter referred to as the Intervention Section, additionally received the sequence of nine structured AI demonstrations described in Section~\ref{sec:intervention} and detailed in Appendix~\ref{apx:demos}; the remaining sections received no AI demonstrations.

\paragraph{Institutional AI tool context.} At the time of this study, Boston University provided enrolled students, faculty, and staff with free access to TerrierGPT, a university-hosted generative AI chat platform that offers access to paid-tier versions of large language models from multiple vendors (including OpenAI, Anthropic, Meta, and Google) through a single password-authenticated web interface. TerrierGPT operates under Boston University's institutional data protection policies, and inputs to the platform are not used to train external models. The platform was publicly available throughout the Spring 2026 semester and was referenced explicitly during several of the AI demonstrations (see Appendix~\ref{apx:demos}). Beyond TerrierGPT, students retained independent access to consumer-facing generative AI tools (e.g., ChatGPT, Claude, Gemini, DeepSeek) through their own personal accounts; the survey instrument did not restrict its definition of ``generative AI tools'' to TerrierGPT and explicitly named these alternatives (see Appendix~\ref{apx:survey}).

\paragraph{Course generative AI policy.} The course-wide syllabus, shared across all three sections, included an explicit generative AI policy reproduced verbatim below:
\begin{quote}
The purpose of this course is to learn new skills. In our opinion, the biggest risk of using generative AI tools for this class is that you will deprive yourself of the opportunity to learn new skills in an environment with many resources to support you. That being said, while you are working outside the classroom, you are permitted to use any resources to which you have access in order to figure out how to solve problems (e.g., using stackoverflow while coding, consulting the course textbook to get more background information). To this end, the core ideological tenet of the course generative AI policy is that ultimately you are responsible for the accuracy and validity of the final products that you create, and you are responsible for what you get out of this course.

For any work conducted outside the classroom (homework, truss project, studying for exams and quizzes) you are permitted, though not necessarily encouraged, to use generative AI tools. For all graded work conducted inside the classroom (quizzes, exams) you are forbidden from using generative AI tools and their use is considered a direct academic conduct violation (i.e., cheating). For all non-graded work conducted inside the classroom (group problem solving) you are encouraged to work collaboratively with other students and solve problems without external resources.

We will conduct multiple informational surveys over the course of the semester to capture data on generative AI use in EK301. We ask that you answer these surveys honestly and thoroughly as they will help us continuously improve the course.
\end{quote}
The intervention examined in this paper layered the AI demonstrations described above on top of this common course-level policy, and was implemented only in the Intervention Section.

\paragraph{IRB approval and consent.} This study was reviewed by the Boston University Charles River Campus Institutional Review Board and determined to qualify for exempt status under 45~CFR~46.104(d)~2(ii) and 3(i)(B)(ii) (Protocol \#8256X, approved December~17, 2025). The approved protocol permitted enrollment of up to 270 participants and included the protocol application, consent script, survey instrument, and recruitment materials. All students enrolled in EK301 in Spring 2026 were eligible to participate. Participation was voluntary, did not affect course grades, and proceeded only after students completed the IRB-approved consent process; only data from consenting students are included in the analyses reported here.

\begin{figure}[h]
    \centering
    \includegraphics[width=\textwidth]{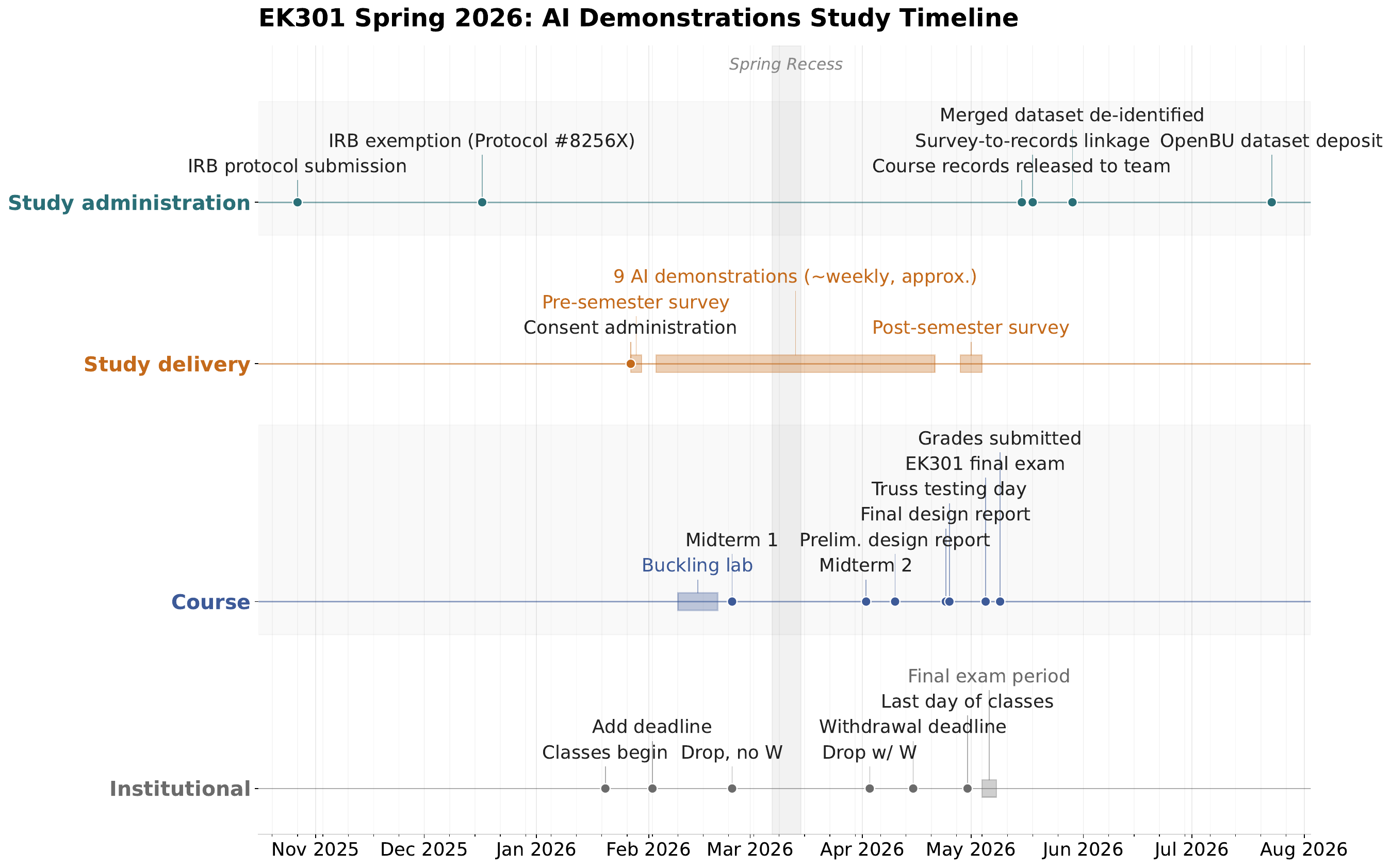}
    \caption{\textbf{Timeline of study execution during the Spring 2026 semester.} Four parallel tracks show study administration milestones (IRB submission, exemption, data linkage, and de-identification), study delivery (consent, pre-semester survey, the nine AI demonstrations delivered at approximately weekly cadence, and post-semester survey), course events (buckling lab,
midterm exams, truss design project milestones, and final exam), and institutional dates (add, drop, and withdrawal deadlines). The shaded band indicates Spring Recess.}    
\label{fig:timeline}
\end{figure}

\subsection{Study Design}
\label{sec:design}

This study employed a quasi-experimental design conducted over the course of a single semester (Spring 2026), integrating three sources of data---a pre-semester student survey, a post-semester student survey, and course assessment records---linked at the individual student level for consenting participants. The pre/post survey structure enabled within-respondent comparison of self-reported AI use, attitudes, and verification practices across the semester. The multi-section structure of EK301 supported a between-sections comparison between students enrolled in the Intervention Section (which received the nine AI demonstrations described in Section~\ref{sec:intervention}) and students enrolled in the remaining sections (which received no AI demonstrations). Section assignment was determined entirely by student registration choices and standard scheduling considerations; students did not select sections based on intervention status and were not informed during registration that one section would receive AI demonstrations. Our intention was that students would not learn which section had been designated as the Intervention Section until after the institutional add deadline (February 2, 2026), and course enrollment records show no evidence of strategic mid-semester section switching in response to anticipated intervention status. Because section assignment was not randomized, and because the analysis does not include covariates that would support adjustment for non-equivalence between groups (see Section~\ref{sec:limitations}), we treat the analyses described in Section~\ref{sec:analysis} as descriptive throughout. Spring 2026 represents the first semester of an ongoing research effort, and the design is expected to evolve across subsequent semesters based on the results reported here and continuing iteration on the demonstration sequence.

The full timeline of study execution is shown in Figure~\ref{fig:timeline}. Briefly, the protocol was reviewed and granted exempt status by the Boston University Charles River Campus IRB in December 2025, prior to the start of the semester (see Section~\ref{sec:course_context} for IRB details). During the second week of the semester, all enrolled students received the IRB-approved consent script and the pre-semester survey, which captured baseline measures of self-rated engineering mechanics understanding, generative AI use frequency, AI-related attitudes and verification practices, and several free-text reflections. Across the remainder of the semester, students in the Intervention Section additionally received the nine AI demonstrations, delivered by the section's regular course instructor at roughly weekly cadence during the final 10--15 minutes of class meetings. All sections completed common course assessments during this period, including weekly homework and quizzes, two midterm exams, and a semester-long truss design project. In the final weeks of the semester, all enrolled students received the post-semester survey, which repeated the core baseline items and added items specific to the AI demonstrations and self-reported changes in AI use over the semester.

After course grades were finalized, course assessment records for consenting students were obtained by the study team under the IRB-approved protocol, merged with consented survey responses at the individual student level, and de-identified to produce the analysis dataset used throughout this paper. All subsequent analyses operated on this merged, de-identified dataset. Details around the public deposition of the de-identified dataset are described in Section~\ref{sec:data_availability}.

\subsection{Study Limitations}
\label{sec:limitations}

This study presents preliminary findings from the first semester of an ongoing research effort, and several limitations should be considered when interpreting the results.  

\noindent \textbf{Study Design and Confounding Variables} \\ 
The overall course enrollment was 196 students, and 105 students consented to have their data used in this study. Coupled with a quasi-experimental design that lacked random assignment, these modest sample sizes preclude strong causal claims. Section assignment was determined by student registration choices and standard scheduling. As a result, the intervention is heavily confounded with other variables, including instructor identity (the intervention section was taught by the Principal Investigator), class meeting times, classroom environments, and unmeasured differences in student composition. Because the intervention was applied to a single section ($n=1$ at the section level), any observed effects cannot be definitively disentangled from these confounding factors. Furthermore, due to FERPA privacy constraints, the analysis does not include covariates for prior academic ability, nor does it report student major, year, or transfer status. As this study was conducted over a single semester at a single institution, the generalizability of the findings remains unknown.  

\noindent \textbf{Measurement and Survey Limitations}\\
Data on student behaviors, attitudes, and AI usage rely entirely on self-reporting, which is susceptible to recall and social desirability biases. The survey lacked baseline measures of students' prior generative AI fluency or out-of-class engagement with traditional course resources (e.g., textbooks, peers). Furthermore, students had varying access to AI tools; while the university provided access to paid-tier models via TerrierGPT, students retained independent access to various consumer-facing platforms, details of which were not tracked. The survey instrument defined generative AI broadly, meaning we cannot control for the highly heterogeneous mix of model capabilities students actually utilized. The AI use frequency measure was recorded on a coarse 5-point ordinal scale, making it impossible to distinguish between a single, prolonged, sophisticated AI session and multiple brief, shallow queries. Survey participation was voluntary but incentivized with a homework bonus point; while this likely boosted overall response rates, it may have encouraged low-effort, brief, or formulaic free-text responses from students primarily motivated by the credit. Additionally, class attendance was generally low and poorly tracked across all sections, and differential survey response rates across sections introduce potential bias into between-section comparisons. A sequencing error in the survey deployment also resulted in one question regarding the impact of the AI demonstrations being mistakenly included in the pre-semester survey before any demonstrations occurred, rendering that specific item uninterpretable. The data also exclude students who dropped or withdrew mid-semester, introducing potential attrition bias if AI use or course difficulty influenced drop decisions. Finally, while a single Likert item was used to check response validity, it remains an imperfect detector of careless or dishonest responding.  

\noindent \textbf{Intervention Delivery and the Nature of LLMs} \\
The structured AI demonstrations evolved dynamically throughout the semester based on instructor judgment rather than following a rigid script. Because the intervention comprised nine distinct demonstrations on different topics, we cannot isolate which specific components drove any observed outcomes. Furthermore, instructor reflections indicate that the timing of certain demonstrations was misaligned with student readiness. For instance, a demonstration on building graphical user interfaces was delivered before students fully grasped the underlying mathematics of the project, and later demonstrations suffered from end-of-semester student fatigue. Moreover, demonstration attendance was self-reported rather than directly observed, which limits any dose-response interpretations within the Intervention Section, and the study lacked granular, continuous feedback mechanisms for individual demonstrations. The instructor reflections provided in Appendix B represent a subjective interpretation of classroom dynamics and should not be over-interpreted. More broadly, generative AI is a rapidly moving target. Tool capabilities, interfaces, and accessibility changed substantively even between January and May 2026. Demonstrations and survey responses anchored to early-semester technology may not reflect the reality of the tools by April, meaning findings from any single semester are unlikely to remain directly applicable in subsequent semesters without ongoing reassessment.  

\noindent \textbf{Analytical and Data Sharing Constraints}\\
The qualitative coding of free-text responses was conducted using a multi-model LLM pipeline. While this pipeline demonstrated high inter-model reliability, it currently lacks a human-coded validation sample; therefore, reported frequencies represent machine-coded categorization (with full audit trails available) rather than human-verified ground truth. Finally, for researchers interested in replicating or extending this work, data sharing is restricted due to de-identification risks regarding sensitive student information. In the version of this dataset that we were able to publish, qualitative and quantitative datasets were required to remain separate, and performance data is limited to broad grade bands rather than specific numerical scores. Our complete and securely held dataset that we are not able to publish due to de-identification risks contains significant additional context that is likely necessary before drawing any conclusions beyond the preliminary ones displayed in this manuscript. \textbf{\textit{We highly recommend that secondary researchers contact the study authors directly for necessary context before attempting data re-use.}}

\subsection{Intervention}
\label{sec:intervention}

Across the Spring 2026 semester, students enrolled in the Intervention Section received a sequence of nine structured AI demonstrations delivered by the course instructor. Each demonstration occupied approximately 10--15 minutes at the end of a regularly scheduled class meeting and recurred at roughly weekly cadence. Demonstrations were designed to be short enough to fit within an existing lecture without displacing core engineering mechanics content, and were sequenced to progress from baseline tool literacy through pedagogical framing, strategic delegation in the context of the truss design project, and finally to model evaluation and benchmarking. Live LLM interactions, when included, were conducted in Google Gemini through the web browser with prior conversation histories cleared before each session to ensure a fresh model context. Students enrolled in all other sections of the course received no AI demonstrations.

Full details of each demonstration, including its pedagogical purpose, the planned content, a summary of any live LLM interactions, and instructor reflections written prior to viewing survey results, are provided in Appendix \ref{apx:demos}. The system prompt used in Demonstration 4 to configure a Socratic tutor is reproduced verbatim in Appendix \ref{apx:prompt}. The nine demonstrations were:
\begin{enumerate}
    \item \textbf{LLM Use in Group Projects:} Demonstrate instructor awareness of current LLM capabilities (including the ability to draft a complete lab report) and set aside dedicated class time for students to discuss what policies around LLM use should govern peer-to-peer communication on group projects.
    \item \textbf{LLMs + The Python Sandbox:} Introduce the concept of a Python sandbox embedded in an LLM interface, contrast sandbox availability across LLM services, and demonstrate iterative use of a sandbox for data visualization.
    \item \textbf{The ``Zone of Proximal Development'':} Introduce ideas from learning science, including metacognition and the concept of a ``More Knowledgeable Other'' \citep{Vygotsky1978}, and facilitate a class discussion of the pros and cons of LLMs serving in that role.
    \item \textbf{LLMs for Learning + System Prompts:} Review three papers from the empirical literature on LLMs and learning \citep{bastani2025generative,team2025ai,shi2025large}, introduce the concept of a system prompt, and demonstrate a custom system prompt designed to convert a general-purpose LLM into an EK301 Socratic tutor (see Appendix \ref{apx:prompt}).
    \item \textbf{Strategic Delegation to LLMs:} Show how students can use LLMs to go beyond baseline project requirements by strategically delegating lower-risk programming tasks, specifically by building a graphical user interface for truss design.
    \item \textbf{Strategic Delegation to LLMs Part II:} Show how students can use LLMs to go beyond baseline project requirements by implementing a simple optimization routine to support iteration on truss design.
    \item \textbf{Benchmarking LLMs:} Introduce the concept of LLM benchmarking, reference popular examples from the literature, and demonstrate an ad hoc EK301-specific benchmark by uploading course problems individually and qualitatively evaluating responses.
    \item \textbf{Benchmarking LLMs Part II:} Demonstrate the structure of a programmatic scientific benchmark, including API calls, multiple model runs, and automated evaluation, drawing from current research and showing how a robust benchmark can be built for EK301 course content.
    \item \textbf{Course Wrap-Up and Student Feedback:} Close the demonstration sequence with a structured reflection activity in which students reviewed all prior demonstrations and provided written and live-polled feedback on their perceived usefulness.
\end{enumerate}

\subsection{Measures}
\label{sec:measures}

This study draws on three sources of data: a pre-semester student survey, a post-semester student survey, and course assessment records collected over the course of the semester. Our goal is to link these sources so that self-reported AI use, attitudes, and behaviors can be examined alongside academic performance in the course.

\subsubsection{Survey Measures}
\label{sec:survey}

Survey data were collected via two Qualtrics instruments: a pre-semester survey administered at the start of the semester and a post-semester survey administered at the end (see Fig. \ref{fig:timeline} for timing). With the exception of items specific to the AI demonstrations and one retrospective item on changes in AI use, all items appeared on both instruments with identical wording, enabling within-respondent comparison of pre- and post-semester responses. The full instrument, including exact item wording, response options, dataset variable names, and notes on items that differed between timepoints, is reproduced in Appendix \ref{apx:survey}.

The survey contained 24 items organized into eight thematic groupings. Section A (four items, 5-point Likert) measured students' self-rated understanding of engineering mechanics, including confidence in applying principles, self-assessed ability to identify and correct mistakes, recognition of the gap between conceptual and procedural understanding, and perceptions of exam scores as a reflection of understanding. Section B (two items, 5-point ordinal) measured self-reported frequency of generative AI use both in general and for engineering mechanics coursework specifically. Section C (five items, 5-point Likert with a Not Applicable option) measured attitudes toward AI tools, including perceived usefulness for coursework, perceived helpfulness for understanding engineering mechanics specifically, whether AI tools introduced confusion, self-rated ability to evaluate the correctness of AI-generated solutions, and concerns about AI accuracy. Section D (two items, 5-point Likert with Not Applicable) measured perceived helpfulness of AI tools relative to two other learning resources --- textbooks and the course discussion section. Section E (one item, 0--10 scale) asked students to rate the overall impact of AI tools on their learning in engineering mechanics.

Section F appeared on the post-semester survey only and captured students' experience with the AI demonstrations: a single item assessing the number of demonstrations attended (integer, 0--9), followed by three Likert/ordinal items on perceived helpfulness, engagement, and influence on AI use, plus one open-ended feedback item. The first item was presented to all post-survey respondents; the four follow-up items were branched only to students who reported attending at least one demonstration. Section G consisted of four free-text reflection items (three asked at both timepoints, one POST-only) on students' typical AI workflow, verification practices, perceived impact of AI tools on their learning, and self-reported changes in AI use over the semester. Section H consisted of a single Likert item at each timepoint asking respondents to indicate whether they had responded honestly and thoughtfully, used as a response-validity check during analysis. One additional item asked at PRE that referenced the AI demonstrations was removed from the deposited dataset because it was administered before any demonstrations had occurred and was judged uninterpretable; this exclusion and the item text are documented in Appendix \ref{apx:survey}.

\subsubsection{Course Performance Measures}

While our initial study design collected a comprehensive set of academic performance metrics, including all individual exam scores, final course grades, prerequisite physics grades, and cumulative GPAs, consultations with the Institutional Review Board (IRB) and strict adherence to the Family Educational Rights and Privacy Act (FERPA) necessitated limiting the published dataset to mitigate the risk of student de-anonymization. 

To evaluate course performance while enabling valid comparisons across the three sections, we selected midterm grades as our primary metric. Unlike final exams and overall course letter grades, which were subject to section-specific curves and grading breakdowns, the two midterm exams were developed collaboratively by all instructors and administered uniformly. These midterms were taken in-class and on paper; students were permitted to use calculators but were strictly prohibited from accessing the internet or external resources. Given the ubiquity of engineering mechanics solutions available online and the ready accessibility of LLMs, we consider performance in this controlled, offline environment to be a much stronger reflection of independent student understanding than out-of-class assignments. 

For our analysis, we calculated each student's average z-score across the two in-class midterms. To further protect student privacy within the public dataset, these standardized scores are reported categorically by grade quartile, where Q1 represents the bottom quartile of performance and Q4 represents the top quartile. In Fig. \ref{fig:timeline}, we show the timing of the midterms with respect to the AI demos. Note that additional AI demos were conducted after the second midterm, which we will aim to address in future study design.

\subsection{Data Preparation, Availability and Ethical Constraints}
\label{sec:data_prep}

Of the 196 students enrolled in the course, 187 unique students completed at least one survey. On paper consent forms were independently coded by two researchers and validated by a third to ensure strict compliance, resulting in 107 consenting participants. Two students were subsequently excluded from this consented pool: one for failing a post-survey honesty check and another for providing clearly invalid responses. This yielded a final analytical sample of $n=105$ consenting students. In cases of repeated survey submissions by a single student, only the first submission was retained to prioritize the most robust qualitative reflections. To ensure data integrity and prevent contamination of the control group, responses from students in the non-intervention sections were carefully screened. Erroneous or ambiguous entries in the demonstration-specific survey columns (e.g., selecting "Neither agree nor disagree" or writing "N/A" for an unattended demonstration) were systematically cleared. A summary of the participant flow is shown in Table \ref{tab:participants}. 

\begin{longtable}{lccc}
\caption{Participant flow and analytic sample composition, EK301 Spring 2026. Section-level enrollment and consent counts are not reported separately because they are identifying; recruitment figures are given as totals only.}
\label{tab:participants} \\
\toprule
 & Intervention & Other & \\
 & section & sections & Total \\
\midrule
\endfirsthead

\multicolumn{4}{l}{\textit{Table \thetable\ continued from previous page}} \\
\toprule
 & Intervention & Other & \\
 & section & sections & Total \\
\midrule
\endhead

\midrule
\multicolumn{4}{r}{\textit{Continued on next page}} \\
\endfoot

\bottomrule
\endlastfoot

\multicolumn{4}{l}{\textit{Recruitment}} \\
Enrolled in EK301                        & --- & --- & 96 \\
Completed at least one survey            & --- & --- & 187 \\
Consented to participate                 & --- & --- & 107 \\
Excluded (response-validity screening)   & --- & --- & 2 \\
\textbf{Final analytic sample}           & \textbf{65} & \textbf{40} & \textbf{105} \\
\midrule
\multicolumn{4}{l}{\textit{Survey completion within the analytic sample}} \\
Pre-semester survey                      & 64 & 36 & 100 \\
Post-semester survey                     & 61 & 38 & 99  \\
Both timepoints                          & 60 & 34 & 94  \\
Pre-semester only                        & 4  & 2  & 6   \\
Post-semester only                       & 1  & 4  & 5   \\
Midterm grade quartile available         & 65 & 40 & 105 \\
\midrule
\multicolumn{4}{l}{\textit{AI demonstration exposure}} \\
Answered the attendance item             & 61 & 38 & 99 \\
Reported attending $\geq$1 demonstration & 60 & 0  & 60 \\
Reported attending all nine              & 28 & 0  & 28 \\
\midrule
\multicolumn{4}{l}{\textit{Grade quartile composition (Q4 = top quartile)}} \\
Q4 & 12 & 14 & 26 \\
Q3 & 18 & 9  & 27 \\
Q2 & 17 & 9  & 26 \\
Q1 & 18 & 8  & 26 \\
\end{longtable}

To comply with institutional privacy policies and the Family Educational Rights and Privacy Act (FERPA), significant steps were taken to mitigate the risk of student de-anonymization. Qualitative and quantitative datasets were completely decoupled and assigned independent, randomized identifiers. Furthermore, all free-text responses were manually redacted to remove potential identifying markers. This included stripping graduation years, removing specific references to peer backgrounds, and replacing exact course prerequisites or personal demographic details with bracketed generic terms (e.g., "[other STEM]" or "[language related difficulties]"). Because of these strict de-identification requirements, the public dataset deposited via OpenBU (\url{https://hdl.handle.net/2144/53355}) separates qualitative text from quantitative performance metrics and reports course performance solely as broad grade quartiles rather than specific numerical scores. \textbf{\textit{Again, we highly recommend that secondary researchers contact the study authors directly for necessary context before attempting data re-use.}}

Dataset variable names for every item follow a consistent convention: variables ending in \texttt{\_pre} originate from the pre-semester survey and variables ending in \texttt{\_post} originate from the post-semester survey. The full variable-to-item mapping is provided in Appendix \ref{apx:survey} and matches the column names in the de-identified dataset deposited via OpenBU. To protect student privacy, the version of the dataset published on OpenBU does not link \textit{qualitative} (i.e., free text) responses to student grades. The dataset contains three files: 
\begin{itemize}
\item \texttt{Spring\_2026\_columnIDs.csv}: A data dictionary containing 46 entries that maps each variable name to its exact survey question text, the corresponding item number in the survey instrument reproduced in Appendix \ref{apx:survey}, and the timepoint (pre or post) at which it was administered. For each variable it also lists the full set of response options in the order presented to participants along with the response type, indicates whether a question was restricted to the demonstration section, and specifies which data file contains the corresponding results.
    \item \texttt{Spring\_2026\_quantitative.csv}: Contains the de-identified structured survey data ($n=105$). This file includes an anonymous \texttt{QuantID}, section assignment status, responses to all ordinal and Likert-scale items across both timepoints, and the student's overall course performance represented as a \texttt{grade\_quartile}.
    \item \texttt{Spring\_2026\_qualitative.csv}: Contains the de-identified free-text survey responses ($n=105$). To prevent the deanonymization of grades based on student writing styles, these responses are keyed to an independent \texttt{QualID}. It captures all open-ended reflections on AI workflow steps, verification habits, perceived impact, and demonstration feedback. 
\end{itemize}
In summary, this tripartite file structure ensures that both structured academic performance data and rich qualitative reflections are accessible for analysis while strictly maintaining student anonymity across the pre- and post-semester timepoints.

\subsection{Analysis}
\label{sec:analysis}

As stated previously, to protect student privacy, quantitative survey responses linked to course performance measures and qualitative responses were reported independently. Details of each analysis approach are described below. 

\subsubsection{Quantitative Results Reporting}
\label{sec:a_quant}

Due to the modest sample size of the dataset ($n=105$), our analysis is strictly descriptive, and we do not report tests of statistical significance or make causal claims. The quantitative results are presented visually to summarize student behaviors, attitudes, and changes over the course of the semester. To visualize the ordinal and Likert-scale data, we employ several specialized plotting strategies. Items measured on a 5-point ordinal frequency scale (e.g., self-reported AI use) are presented as stacked horizontal bar charts. These are plotted left-aligned, with percentages displayed inline for segments exceeding 5\% of the total. 

For 5-point Likert-scale items measuring agreement or attitudes, we utilize diverging stacked bar charts. These plots anchor the neutral response ("Neither Agree nor Disagree") squarely on a center zero-line. Negative sentiments (Disagree/Strongly Disagree) expand to the left, and positive sentiments (Agree/Strongly Agree) expand to the right. This diverging format allows for an immediate, intuitive visual comparison of net sentiment shifts between the pre-semester and post-semester timepoints. The raw paired rating of the impact of AI tools (measured on a 0-10 scale) is reported using paired, stacked histograms denoting the mean shift. Furthermore, to explore the relationship between academic performance and AI tool use without deanonymizing students, we cross-tabulate self-reported AI frequency against course performance. This is visualized using a faceted stacked bar chart broken down by student grade quartiles (Q1-Q4), representing the standardized average of the two in-class midterms.

To ensure full transparency and reproducibility, all Python scripts used to process the dataset and generate these visualizations, including the central \texttt{plot\_style.py} configuration, the custom \texttt{load\_research\_data.py} parser, and the plotting logic for both quantitative and qualitative responses, are publicly available in the project's GitHub repository.

\subsubsection{Qualitative coding of free-text responses}
\label{sec:a_qual}

Free-text responses (Items 19--23 of the survey instrument, see Appendix \ref{apx:survey}) were coded against structured codebooks developed by the research team (see Appendix \ref{apx:codebook}). The codebooks were authored by members of the research team, who drafted the initial categories from a full reading of the response corpus and revised them through group discussion. Because the corpus spans eight question-timepoint combinations and 756 response-question pairs, we used a multi-model machine coding pipeline with human-auditable outputs rather than traditional manual coding, following an approach we document in full in Appendix \ref{apx:llm_coding}, to match each student response to codebook entries. Briefly, each response was coded independently by three LLMs from three different providers (Claude Sonnet 5, GPT-5.5, and Gemini 3.5 Flash), each receiving the survey question, the complete codebook with decision rules and example responses, and the response text, and returning structured output that was programmatically validated against the codebook's constraints (e.g., multiple codes per response where permitted, exactly one Tier 1 code for tiered codebooks, and standalone codes never combined with others). The three LLM coders never saw one another's output, respondent identity, or section assignment.

Responses that received identical code sets from all three LLM coders (445 of 756 response-question pairs, 58.9\%) were accepted directly. The remainder were resolved by an adjudication step in which a fourth model configuration (Claude Fable 5) received the response, the codebook, and all three codings with their rationales, and assigned final codes judging strictly against the codebook rather than by majority vote. Overall, inter-coder reliability was high. Specifically, across the 95 codes with mean prevalence of at least 5\%, median Fleiss' $\kappa$ was 0.88 (IQR 0.78--0.94) and 90 of 95 (94.7\%) reached the conventional threshold for substantial agreement ($\kappa \geq .61$) \citep{fleiss1971measuring,landis1977measurement}; mean pairwise Jaccard similarity of code sets ranged from 0.76 to 0.92 across the eight question-timepoint combinations (Appendix \ref{apx:llm_coding}, Table~\ref{tab:llm_reliability}). We emphasize that inter-model agreement establishes the reliability of the machine coding, not its validity against human judgment. We adopt this machine-coded, human-auditable approach to disseminate results to other educators quickly under funding and personnel constraints. All code required to reproduce the qualitative coding is published on GitHub, and the coding results are used solely for data visualization. A detailed comparison of this approach against traditional human coding is left outside the scope of this study. All coded frequencies reported in Section \ref{sec:res} are drawn from the final adjudicated coding. Blank responses were treated as item nonresponse and excluded from coding denominators; typed non-answers (e.g., ``NA'') were retained and coded to each codebook's Uncodable category.

\section{Results and Discussion}
\label{sec:res}

We organize the results around the three research questions. All results are descriptive; percentages refer to survey respondents at the indicated timepoint (pre: $n = 100$; post: $n = 99$; demonstration feedback: $n = 60$ demo-section respondents), and coded frequencies are drawn from the final adjudicated qualitative coding described in Section~\ref{sec:analysis}.

\subsection{How Do Students Use LLMs in Engineering Mechanics?}

\begin{figure}[p]
    \centering
    \includegraphics[width=.9\textwidth]{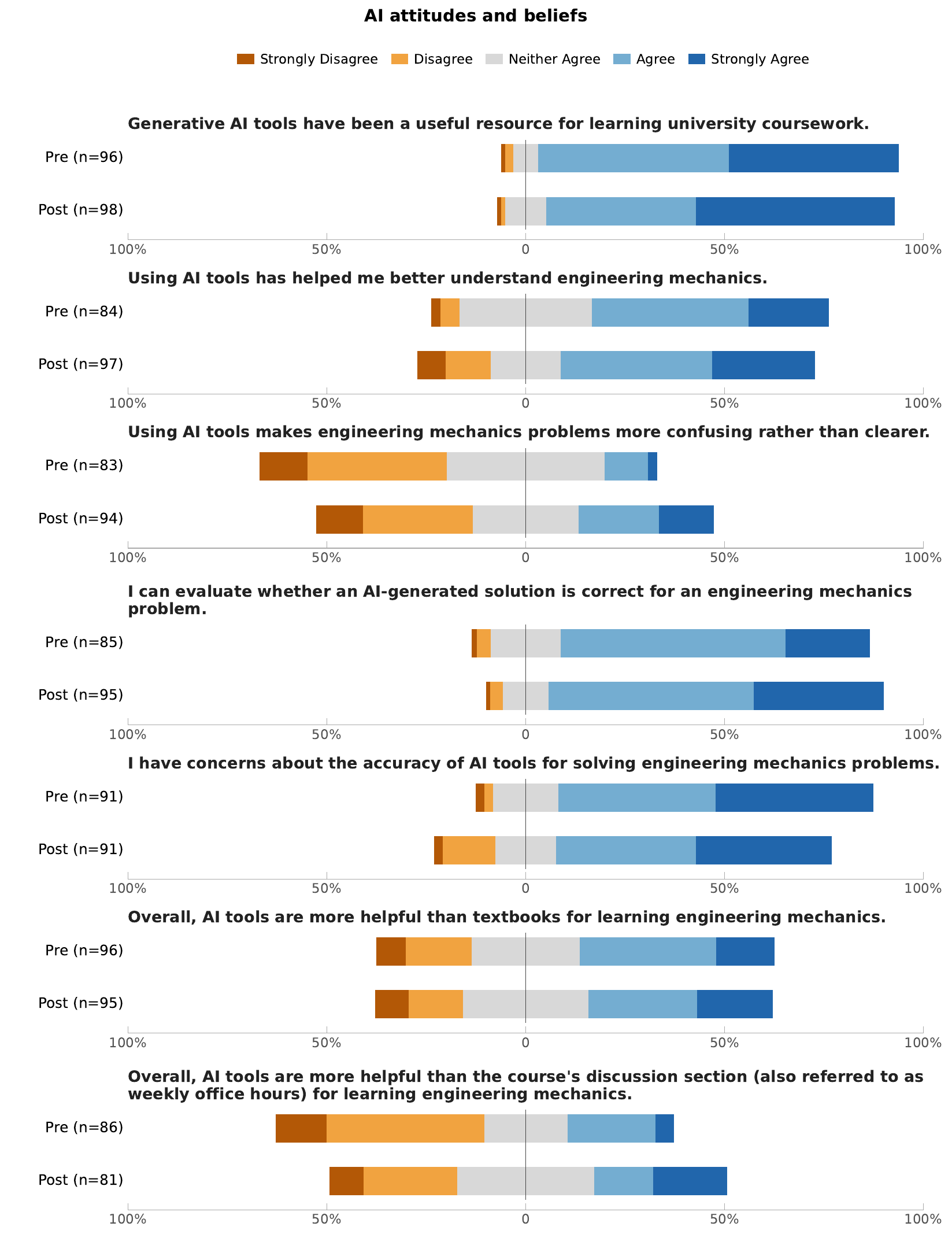}
	\caption{\textbf{Student attitudes toward generative AI tools and AI relative to other learning resources, pre- versus post-semester.} Diverging stacked bar charts of responses to seven 5-point Likert items (Strongly Disagree to Strongly Agree), with the neutral category centered on zero; disagreement extends left and agreement extends right. Each item is shown at the pre-semester (Pre) and post-semester (Post) timepoints, with per-item respondent counts indicated on the axis labels. Respondents who selected Not Applicable are excluded from the corresponding item.}     
	\label{fig:attitudes}
\end{figure}

\begin{figure}[p]
    \centering
    \includegraphics[width=.8\textwidth]{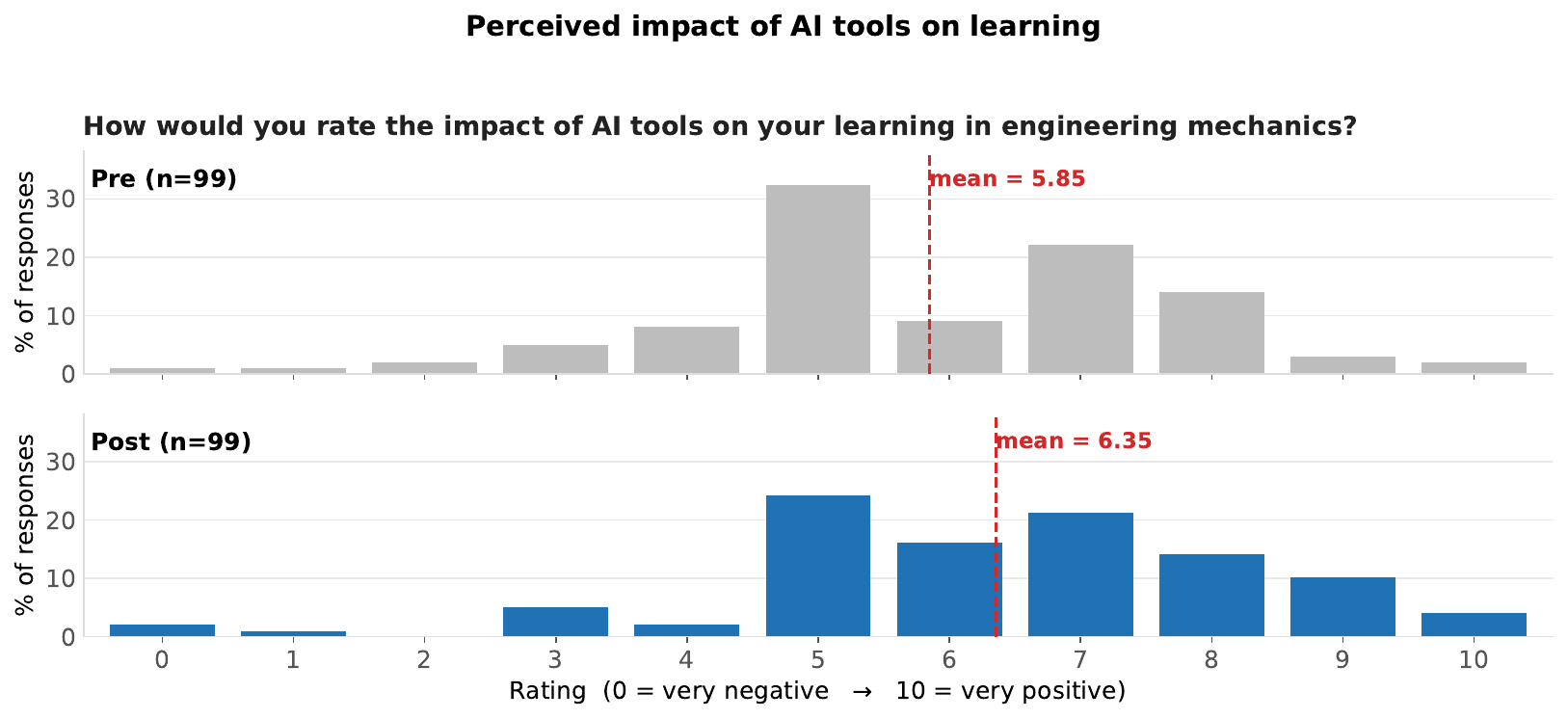}    
    
    \vspace{5mm}
    
    \includegraphics[width=.85\textwidth]{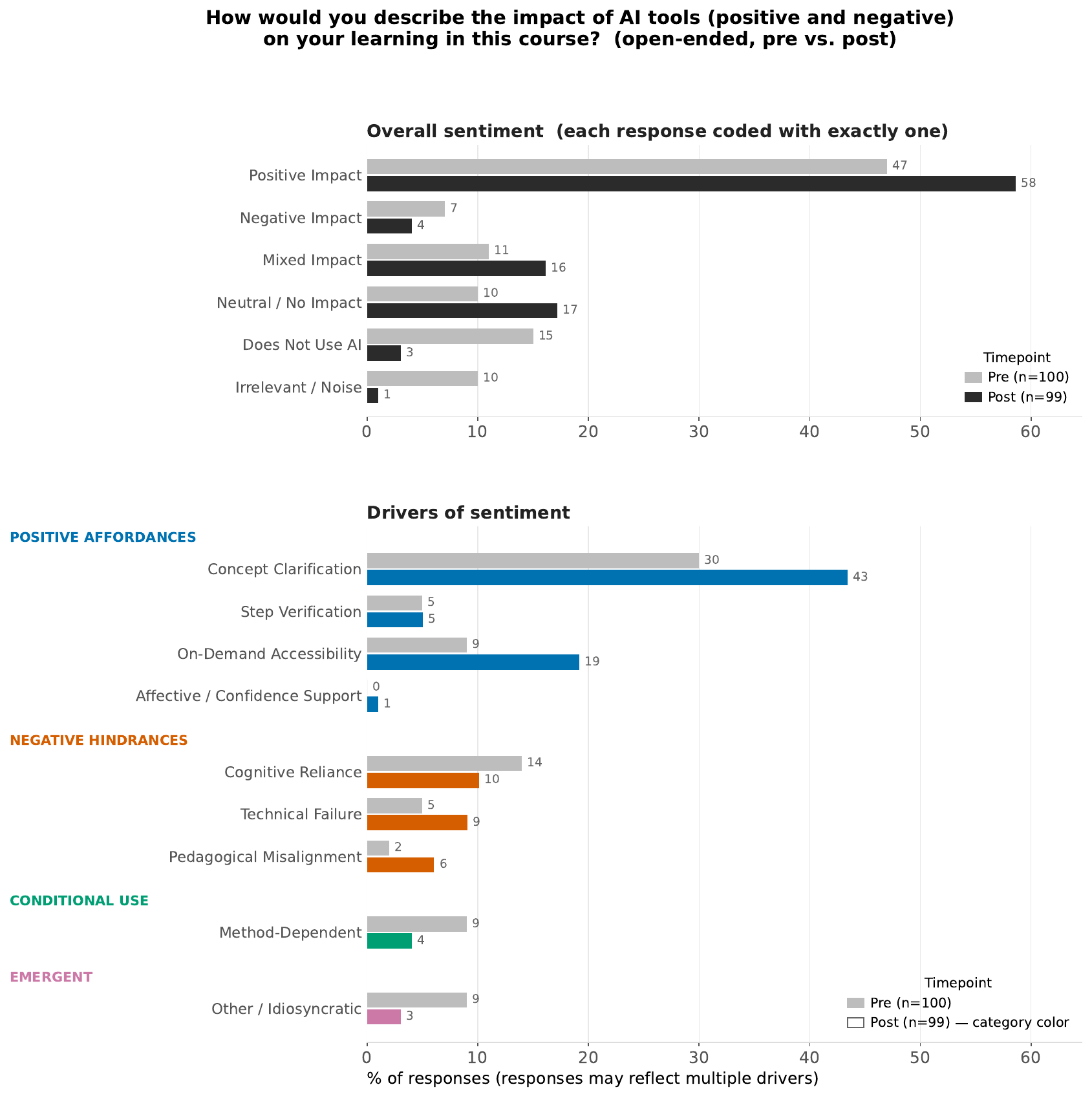}
\caption{\textbf{Perceived impact of AI tools on learning in engineering mechanics, pre- versus post-semester.} Top: distributions of responses to the item asking students to rate the impact of AI tools on their learning, on a 0 (very negative) to 10 (very positive) scale (n = 99 at each timepoint); dashed lines mark the means (5.85 pre, 6.35 post). Bottom: coded free-text responses describing the impact of AI tools on learning (pre n = 100, post n = 99). Overall sentiment was coded with exactly one code per response; drivers of sentiment could receive multiple codes per response. Frequencies are drawn from the final adjudicated machine coding described in Section \ref{sec:a_qual}.}
\label{fig:impact}
\end{figure}

\begin{figure}[p]
    \centering
    \includegraphics[width=\textwidth]{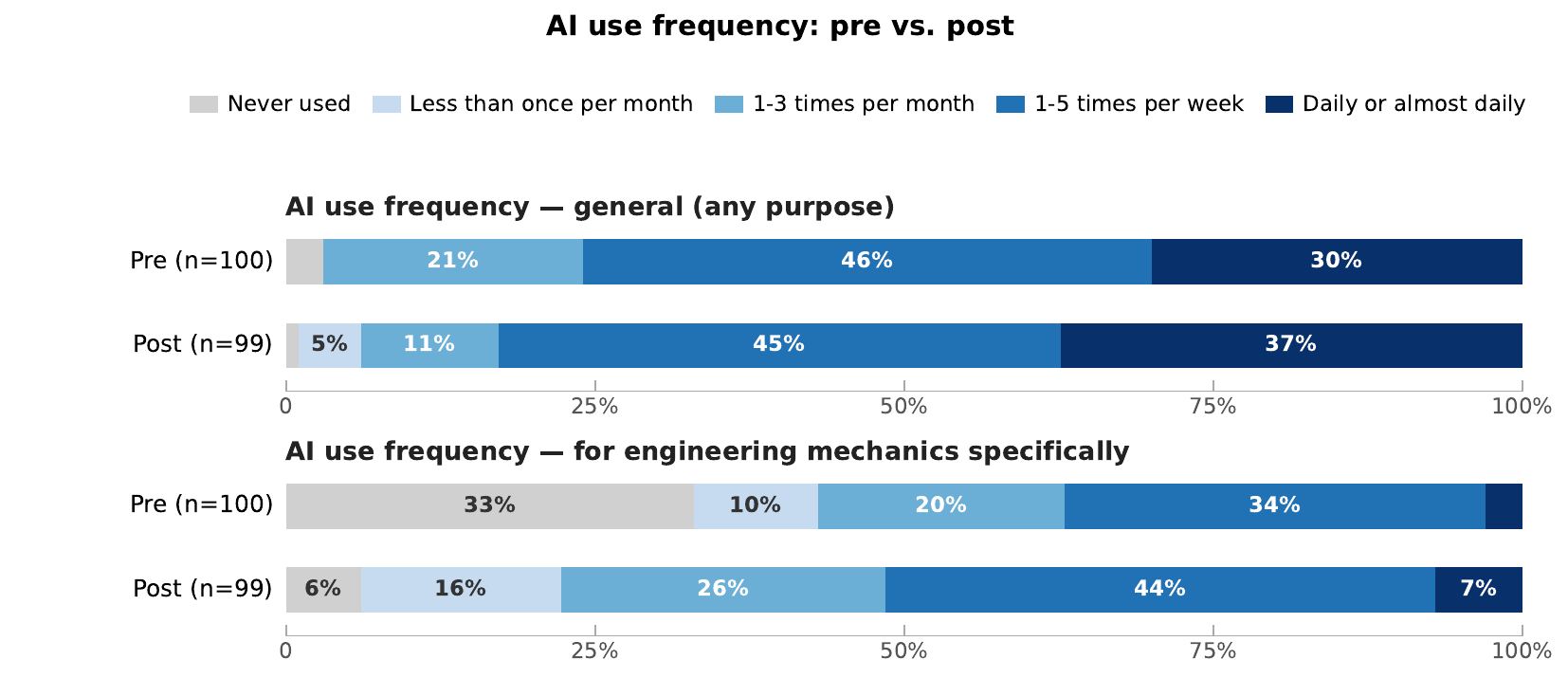}    
    
    \vspace{5mm}
    
    \includegraphics[width=.9\textwidth]{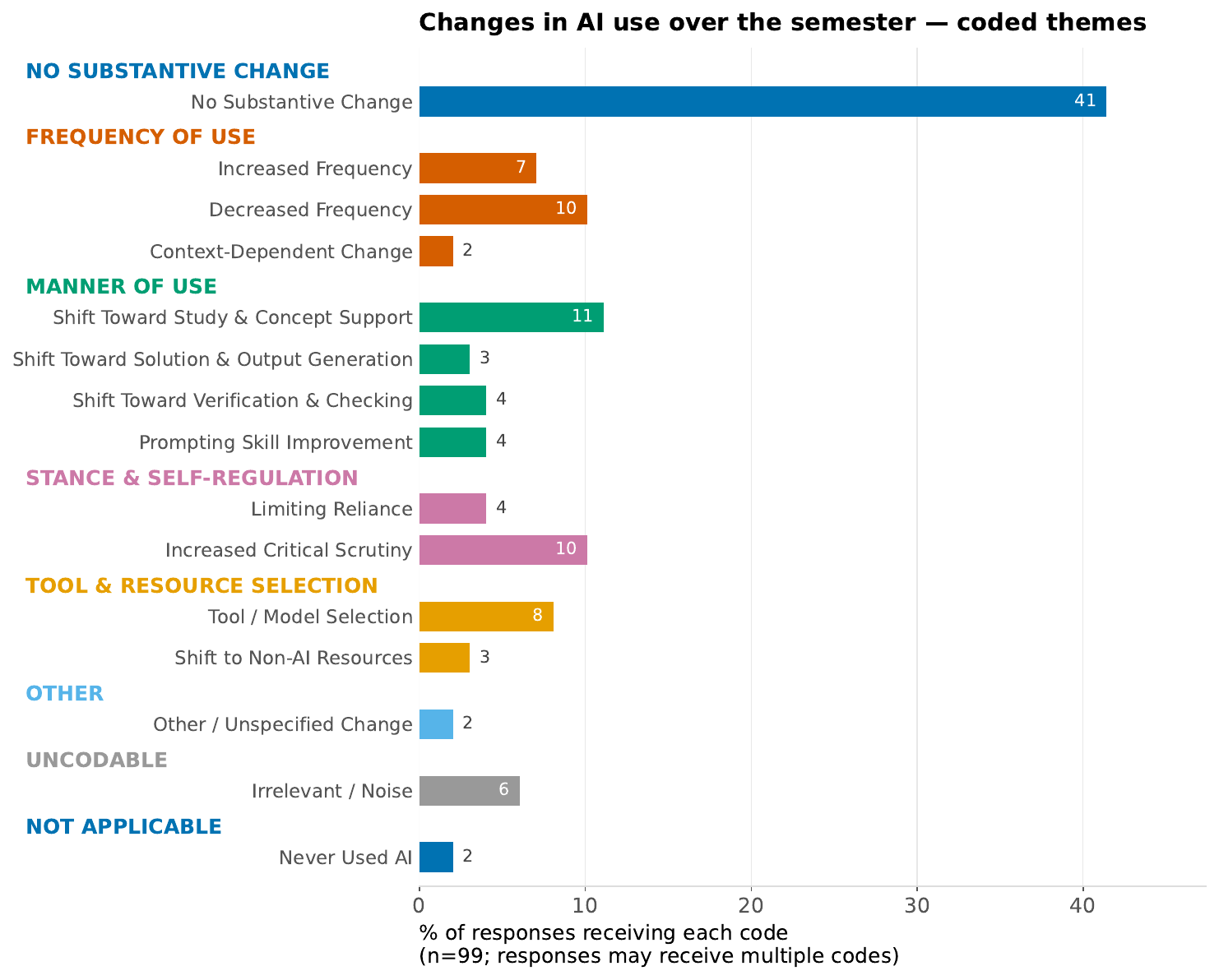}
\caption{\textbf{Self-reported AI use frequency and coded changes in AI use over the semester.} Top: stacked horizontal bar charts of self-reported generative AI use frequency on a 5-point ordinal scale (Never used to Daily or almost daily), for general use (any purpose) and for engineering mechanics coursework specifically, at the pre-semester (n = 100) and post-semester (n = 99)
timepoints; inline percentages are shown for segments exceeding 5\%. Bottom: coded free-text responses to the post-only item asking students to describe changes in how they use AI tools since the start of the semester (n = 99); responses could receive multiple codes, and frequencies are drawn from the final adjudicated machine coding described in Section \ref{sec:a_qual}.}    \label{fig:use_change}
\end{figure}

Generative AI use was near-universal for general purposes and expanded substantially within engineering mechanics over the semester (Fig.~\ref{fig:use_change}, top). At baseline, 76\% of respondents reported using AI tools at least weekly for some purpose, rising to 82\% at post. For engineering mechanics specifically, the shift was larger: the share who had never used AI for mechanics coursework fell from 33\% at pre to 6\% at post, and at-least-weekly use rose from 37\% to 51\%. Attitudes were broadly positive and stable (Fig.~\ref{fig:attitudes}): large majorities at both timepoints agreed that AI tools are a useful resource for university coursework and that they can evaluate whether an AI-generated solution is correct, while concerns about accuracy for mechanics problems remained widespread at both timepoints. Students rated AI tools slightly more helpful than textbooks, and at the pre semester they found them less helpful than the course's discussion section though by the end of the semester they were rated similarly. Consistent with these attitudes, the mean self-rated impact of AI on learning rose from 5.85 to 6.35 on the 0--10 scale, with the post distribution shifting toward the positive end (Fig.~\ref{fig:impact}, top).

\begin{figure}[p]
    \centering
    \includegraphics[width=.9\textwidth]{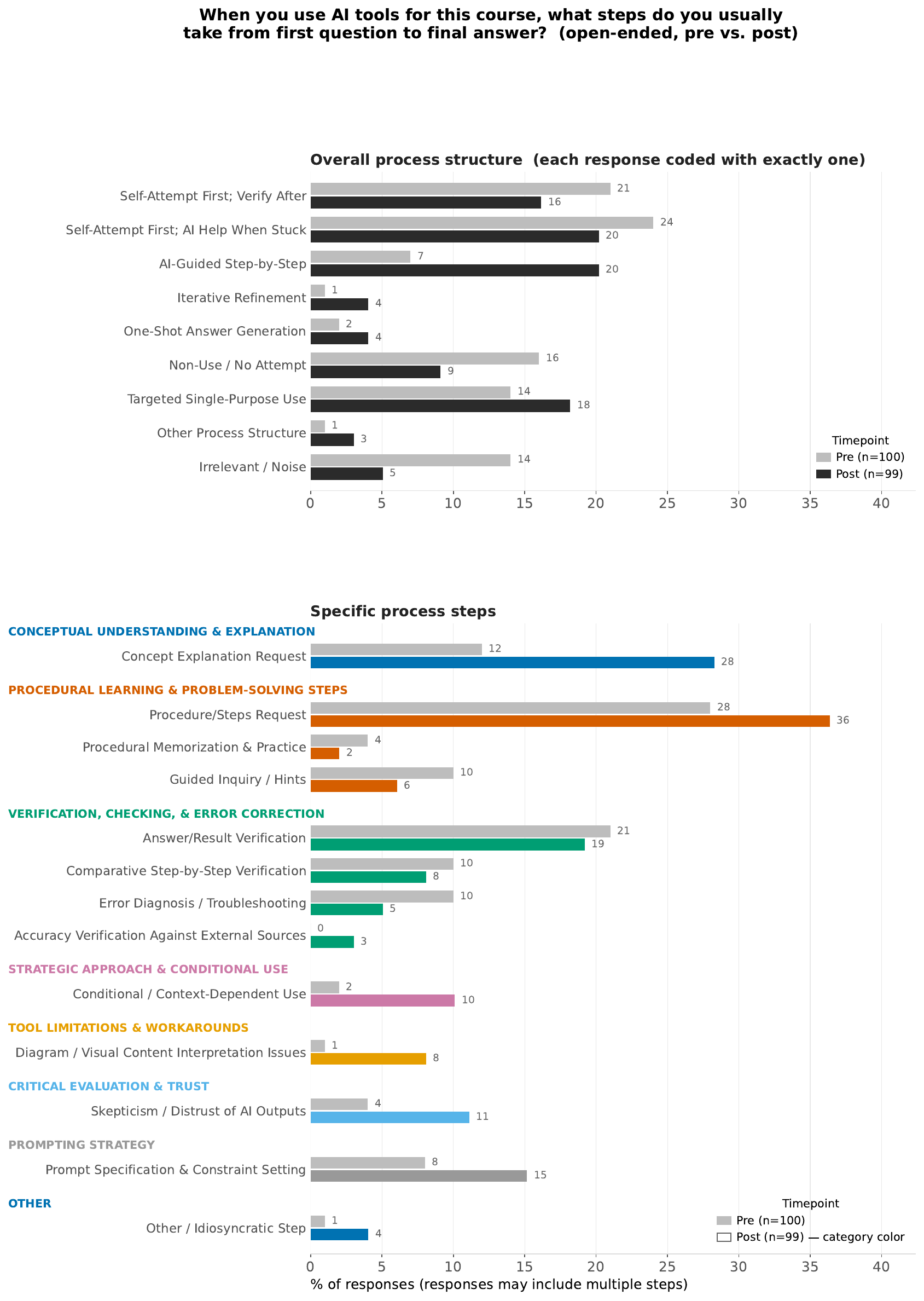}        
   \caption{\textbf{Coded free-text descriptions of students' AI workflows, pre- versus post-semester.} Responses to the open-ended item asking what steps students usually take from first question to final answer when using AI tools (pre n = 100, post n = 99). Top: overall process structure, where each response received exactly one mutually exclusive code. Bottom: specific process steps grouped by thematic category, where responses could receive multiple codes. Bars show the percentage of responses receiving each code at each timepoint; frequencies are drawn from the final adjudicated machine coding described in Section \ref{sec:a_qual}.}
\label{fig:ai_workflow}
\end{figure}

The coded free-text responses add texture to these ratings. Overall sentiment about AI's impact shifted positive (47 to 58 Positive Impact codes pre to post) while non-use nearly vanished (15 to 3 Does Not Use AI codes), and the dominant stated driver at both timepoints was concept clarification (30 to 43 mentions), followed at post by on-demand accessibility (9 to 19) (Fig.~\ref{fig:impact}, bottom). Negative drivers were less common and mixed in direction: mentions of cognitive reliance declined (14 to 10) while mentions of technical failure and pedagogically misaligned explanations increased (5 to 9 and 2 to 6, respectively). Coded workflows (Fig.~\ref{fig:ai_workflow}) tell a parallel story. Self-attempt-first structures, in which students work independently and use AI either to check finished work or to get unstuck, were the most common overall process at both timepoints (45 of 100 pre, 36 of 99 post, combining the two self-attempt codes), but AI-guided step-by-step workflows nearly tripled (7 to 20) as non-use fell (16 to 9). Among specific steps, concept explanation requests more than doubled (12 to 28), procedure and formula requests rose (28 to 36), and three sophistication markers grew: deliberate prompt constraint setting such as instructing the tool not to reveal the answer (8 to 15), explicit skepticism about AI output (4 to 11), and awareness of diagram-interpretation failures (1 to 8).

\begin{figure}[p]
    \centering
    \includegraphics[width=\textwidth]{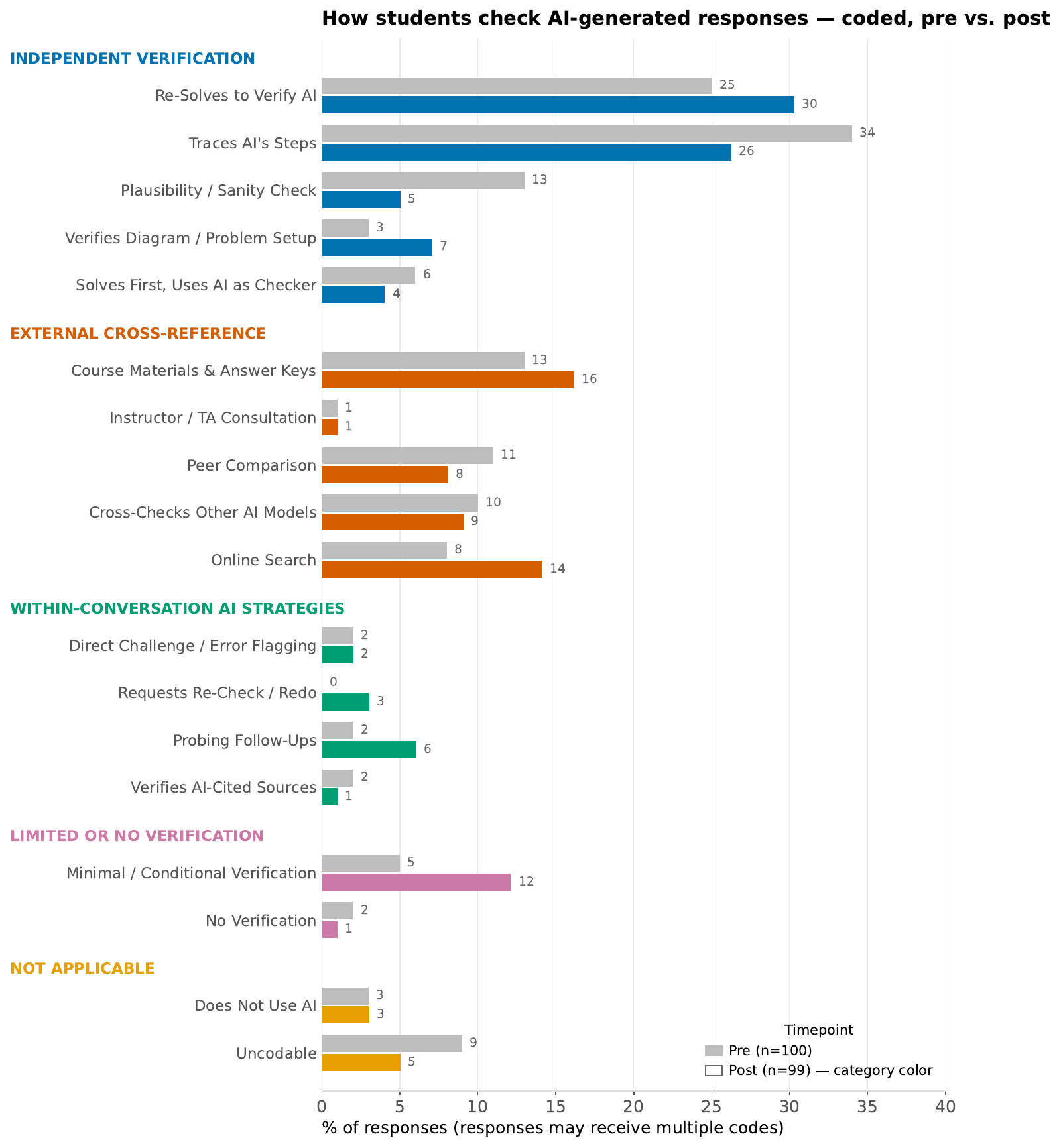}        
	\caption{\textbf{Coded free-text descriptions of how students verify AI-generated responses, pre- versus post-semester.} Responses to the open-ended item asking what, if anything, students do to check whether an AI-generated response is correct (pre n = 100, post n = 99), grouped into five thematic categories: independent verification, external cross-reference, within-conversation AI strategies, limited or no verification, and not applicable. Bars show the percentage of responses receiving each code at each timepoint; responses could receive multiple codes, and frequencies are drawn from the final adjudicated machine coding described in Section \ref{sec:a_qual}.}
    \label{fig:verify_correct}
\end{figure}

Verification behavior shifted more ambiguously (Fig.~\ref{fig:verify_correct}). Independent verification remained the modal strategy, with re-solving to check the AI rising (25 to 30) while passive tracing of the AI's steps fell (34 to 26), and external cross-referencing against course materials and online sources grew (13 to 16 and 8 to 14). At the same time, responses describing only minimal or conditional verification more than doubled (5 to 12), so increased use did not uniformly translate into increased checking. That checking did not keep pace with use suggests verification is a habit instructors will need to scaffold deliberately rather than assume, especially as attention turns to how LLM use can be proactively guided \citep{geng2026}. Finally, when asked directly what had changed (Fig.~\ref{fig:use_change}, bottom), the single most common response was no substantive change (41 of 99), and among reported changes, decreases in frequency slightly outnumbered increases (10 to 7) while the most common substantive shifts were toward study and concept support (11) and increased critical scrutiny (10). The contrast between this plurality perception of stability and the clear pre to post movement in the distributions above suggests that it's possible that much of the aggregate change was gradual enough to escape students' own retrospective notice, occurred among students who were light users at baseline, or both.

\begin{figure}[p]
    \centering
    \includegraphics[width=\textwidth]{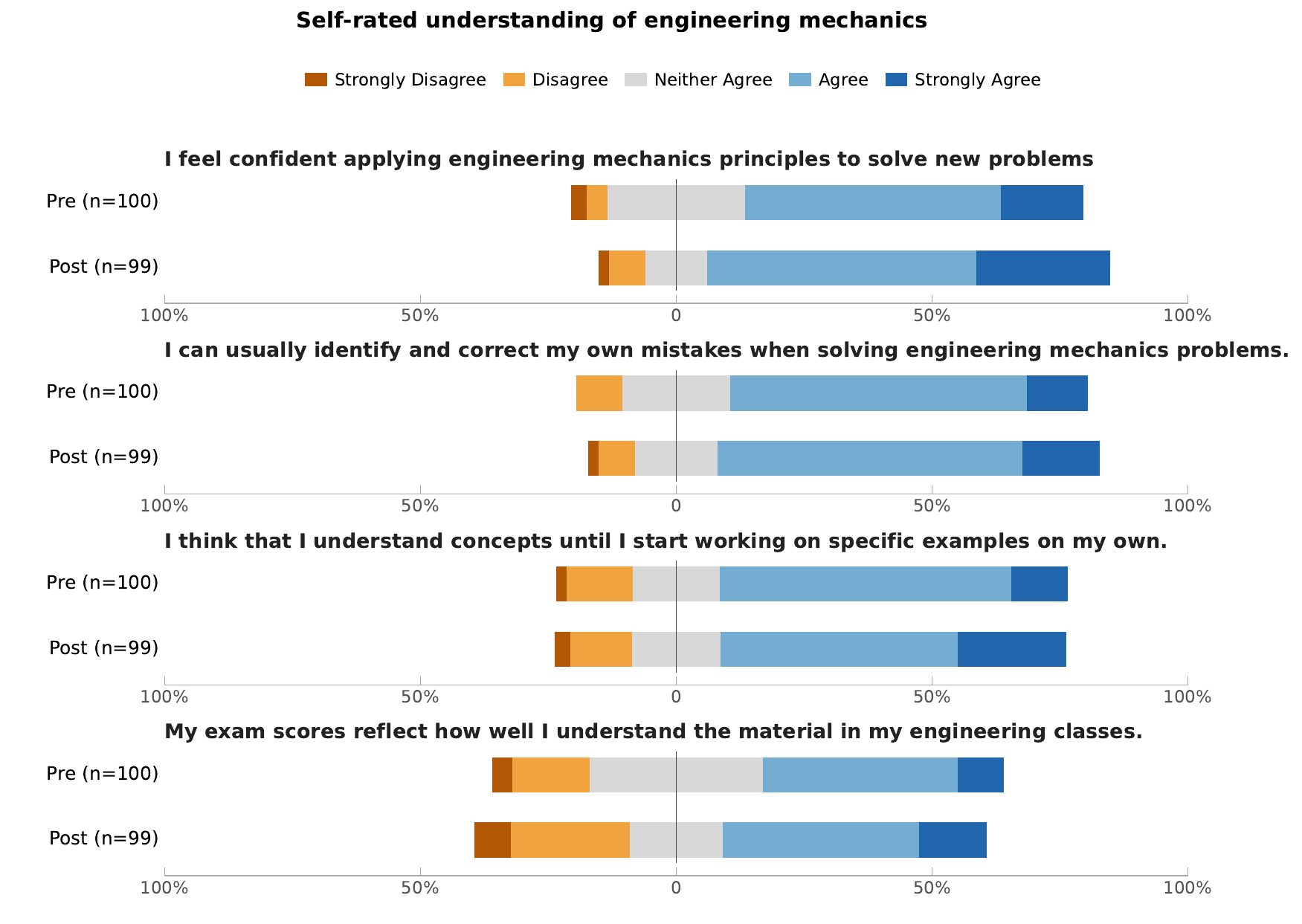}
\caption{\textbf{Self-rated understanding of engineering mechanics, pre- versus post-semester.} Diverging stacked bar charts of responses to four 5-point Likert items (Strongly Disagree to Strongly Agree) measuring confidence in applying engineering mechanics principles, ability to identify and correct one's own mistakes, the gap between conceptual and procedural understanding, and perceptions of exam scores as a reflection of understanding (n = 100 pre, n = 99 post). The neutral category is centered on zero, with disagreement extending left and agreement extending right.}    
\label{fig:understanding}
\end{figure}

\begin{figure}[p]
    \centering
    \includegraphics[width=\textwidth]{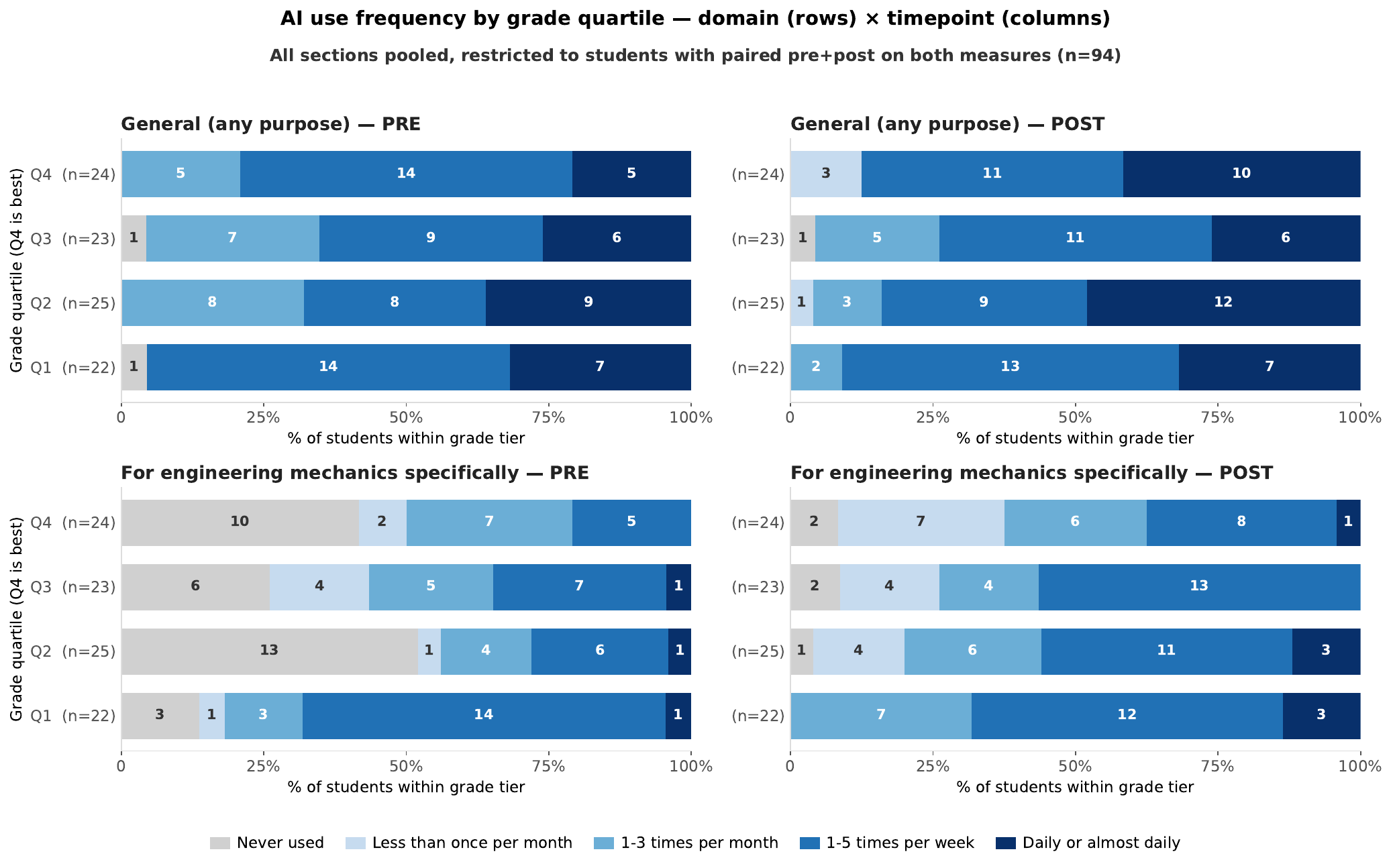}
\caption{\textbf{Self-reported AI use frequency by course performance quartile.} Stacked bar charts of AI use frequency (5-point ordinal scale, Never used to Daily or almost daily) cross-tabulated by grade quartile, with domain in rows (general use versus engineering mechanics specifically) and timepoint in columns (pre versus post). Grade quartiles are derived from each student's average z-score across the two common in-class midterm exams, pooled across all sections, with Q4 the top quartile. The sample is restricted to students with paired pre and post responses on both frequency measures (n = 94); counts within each bar segment are shown.}
    \label{fig:performance}
\end{figure}

\subsection{How Does LLM Use Relate to Course Performance?}

Self-rated understanding of engineering mechanics rose modestly over the semester (Fig.~\ref{fig:understanding}). Specifically, agreement with confidence in applying principles and in identifying one's own mistakes both increased, as expected across a semester of instruction, while agreement that exam scores reflect understanding declined at post. The relationship between AI
use and course performance differed sharply by domain (Fig.~\ref{fig:performance}). General-purpose AI use frequency was high in every grade quartile at both timepoints, with no clear performance gradient. Mechanics-specific use at baseline, however, was inversely related to performance: 10 of 24 top-quartile students (Q4) reported never using AI for mechanics at pre, against 3 of 22 in the bottom quartile (Q1), where 14 of 22 already reported weekly use. By post, never-use had largely disappeared in all quartiles and the gradient narrowed, though Q4 remained the lightest-using group. We emphasize that these are cross-sectional descriptive patterns. Critically, they are equally consistent with struggling students turning to AI for help and with heavier AI reliance displacing independent practice, and this design cannot distinguish those directions.

\begin{figure}[p]
    \centering
    \includegraphics[width=.8\textwidth]{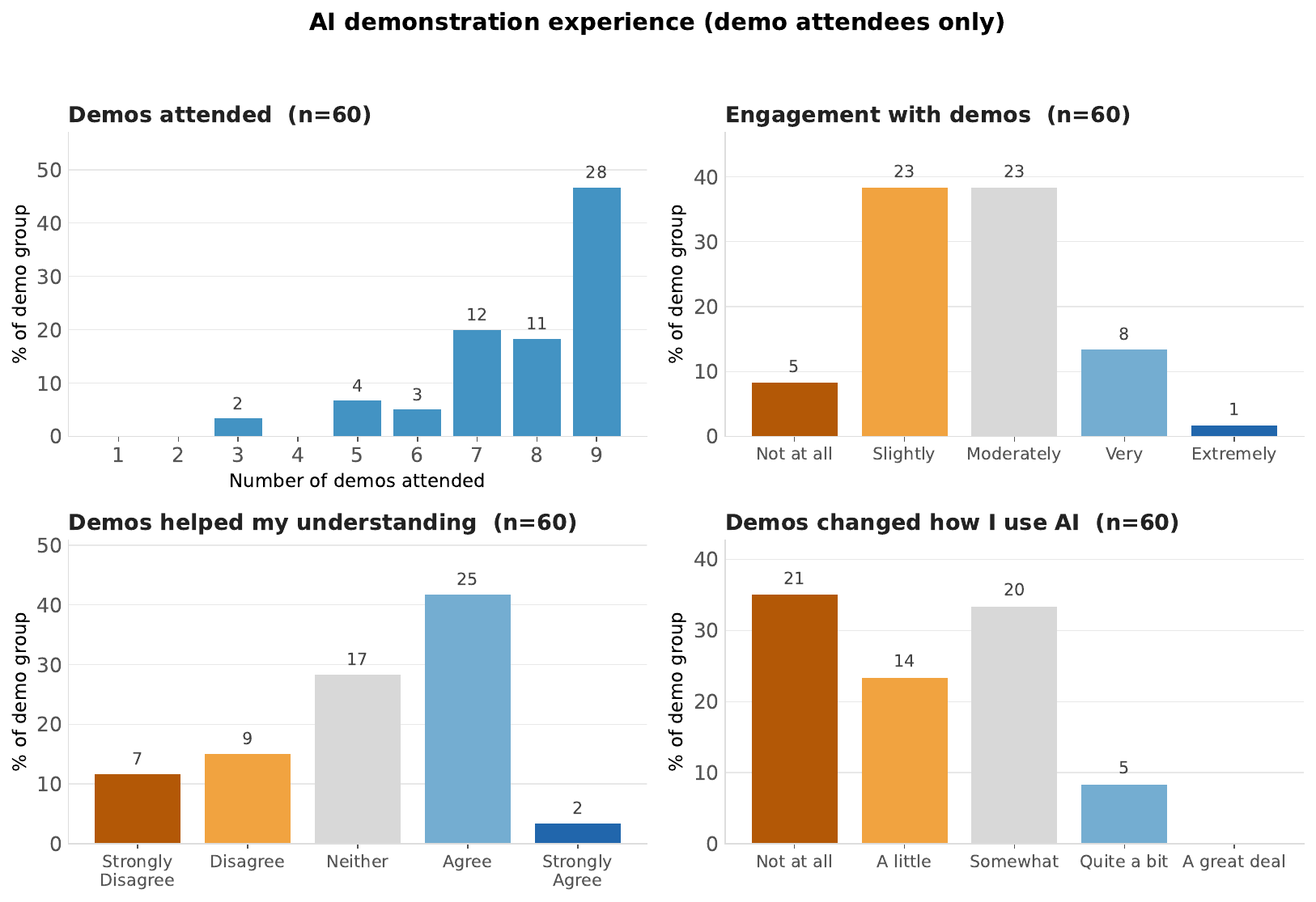}
     \includegraphics[width=.8\textwidth]{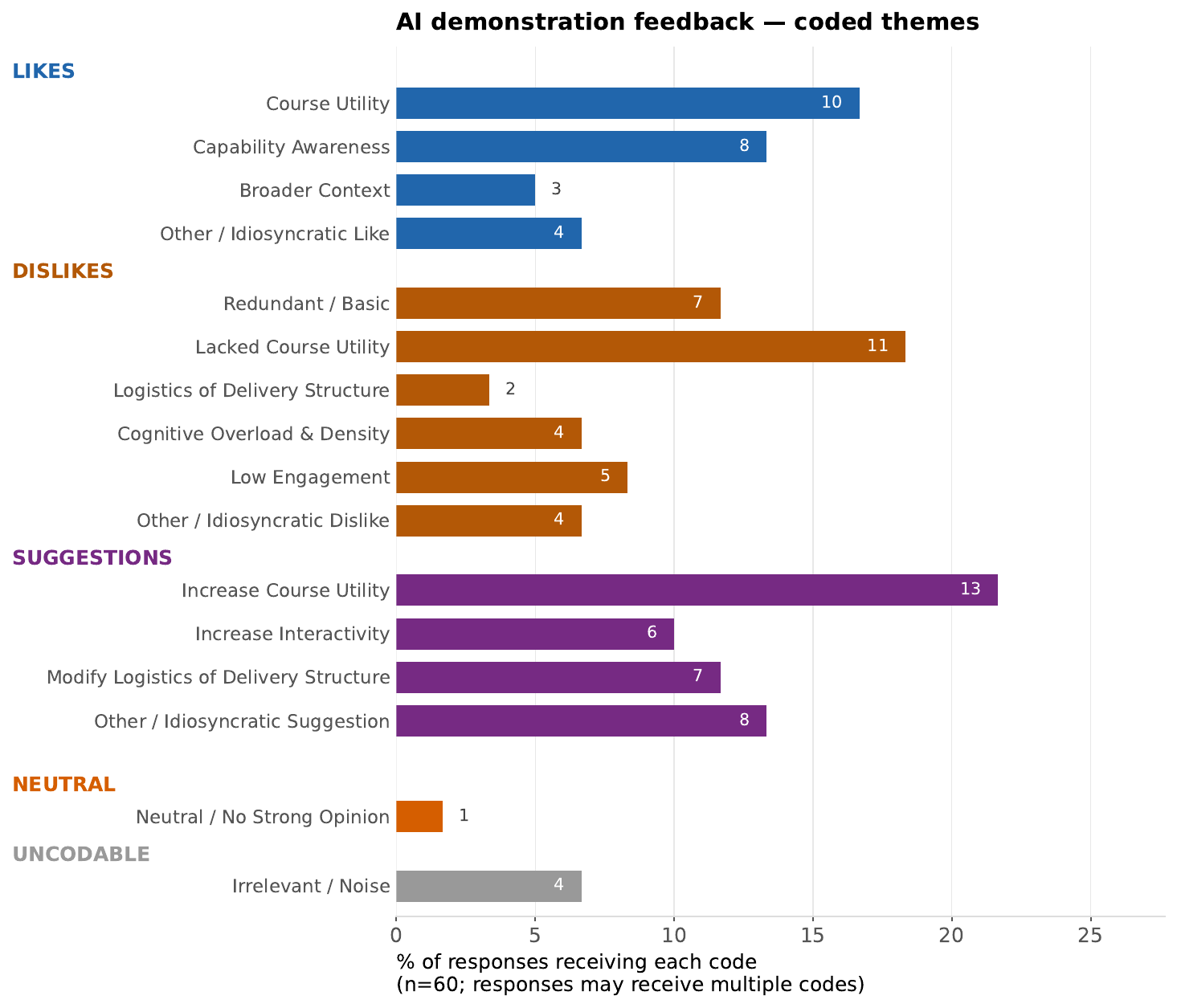}
\caption{\textbf{Student experience with and feedback on the structured AI demonstrations, Intervention Section respondents only.} Top: distributions of self-reported number of demonstrations attended, engagement, agreement that the demonstrations helped strengthen understanding, and the extent to which the demonstrations changed AI use (n = 60 respondents reporting attendance of at least one demonstration). Bottom: coded free-text feedback on the demonstrations grouped into likes, dislikes, suggestions, neutral, and uncodable categories; responses could receive multiple codes, and frequencies are drawn from the final adjudicated machine coding described in Section \ref{sec:a_qual}.}    
\label{fig:demo_feedback}
\end{figure}

\subsection{What Was the Impact of the Structured AI Demonstrations?}

Among the 60 demo-section respondents, self-reported attendance was high (28 reported attending all nine demonstrations) but engagement was modest. Most students described themselves as slightly or moderately engaged (23 each), with only 9 very or extremely engaged (Fig.~\ref{fig:demo_feedback}, top). Twenty-seven of 60 agreed or strongly agreed that the demonstrations helped strengthen their understanding, and 25 reported that the demonstrations changed how they use AI at least somewhat, while 21 reported no change at all. The coded open-ended feedback was sharply split on a single axis: course utility (Fig.~\ref{fig:demo_feedback}, bottom). The most frequent like was direct course utility (10) and the most frequent dislike was lacked course utility (11), with redundancy for students who already knew the material close behind (7); accordingly, the dominant suggestion was to increase course utility (13), followed by modifying delivery logistics (7) and adding interactivity (6). This polarization is consistent with the instructor reflections in Appendix~\ref{apx:demos}. The same demonstrations landed very differently across a cohort with heterogeneous prior AI fluency, and the clearest student mandate for future iterations is tighter coupling between the demonstrations and graded coursework.

\begin{figure}[p]
    \centering
     \includegraphics[width=.9\textwidth]{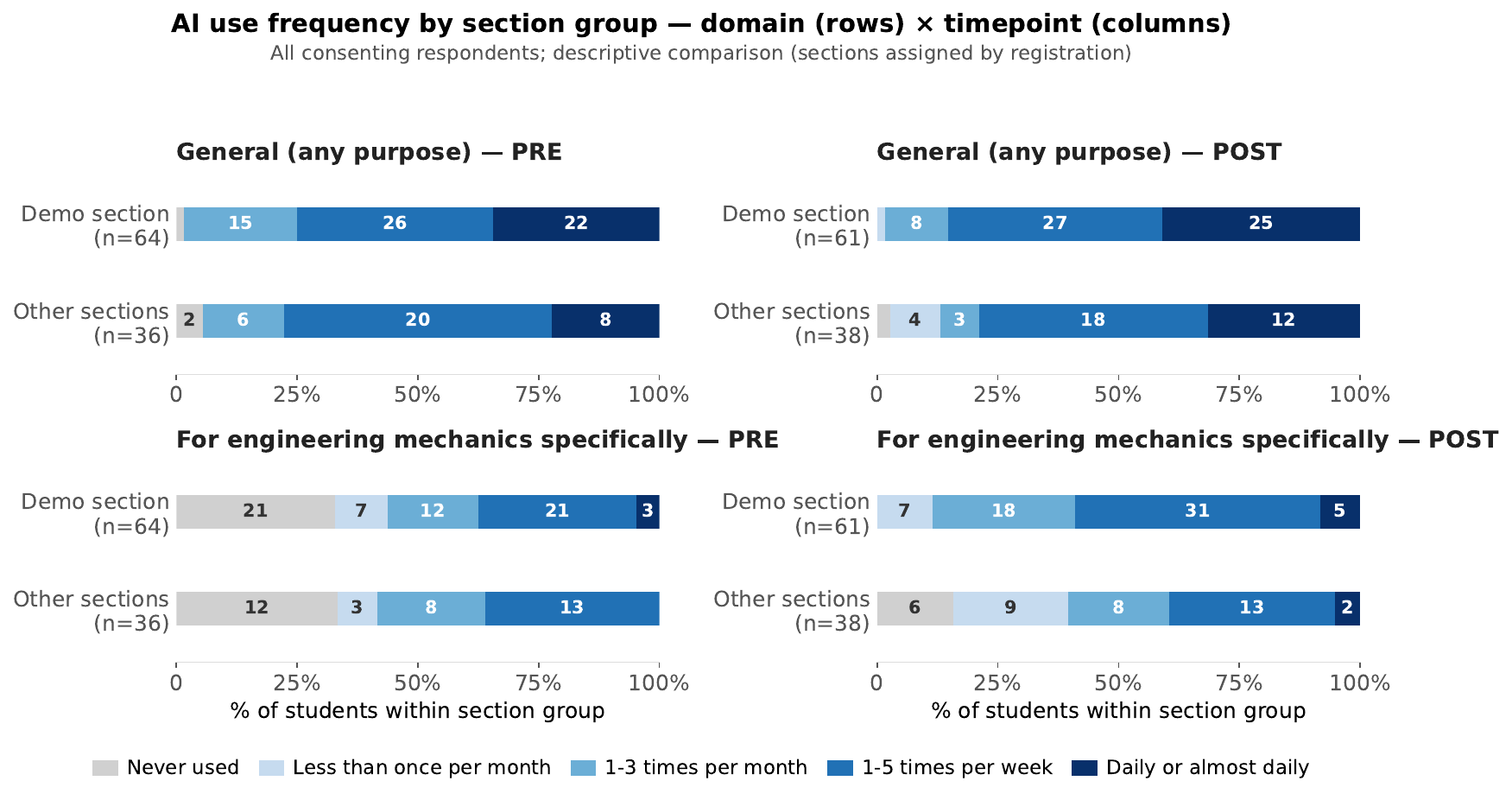}
    
    \vspace{15mm} 
    
     \includegraphics[width=.9\textwidth]{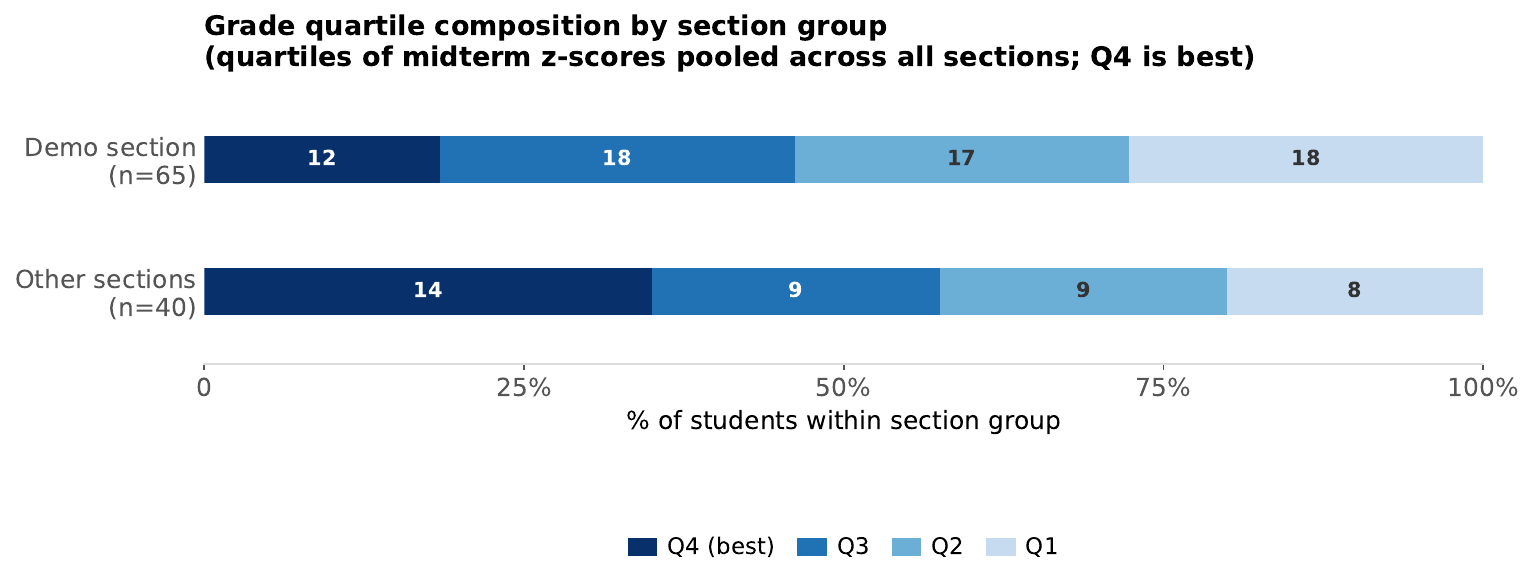}\\

\caption{\textbf{AI use frequency and grade quartile composition by section group.} Top: stacked bar charts of self-reported AI use frequency (5-point ordinal scale) by section group (Intervention Section versus all other sections), with domain in rows and timepoint in columns; all consenting respondents are included, and sections were assigned by student registration rather than randomization. Bottom: grade quartile composition of each section group, where quartiles are derived from midterm z-scores pooled across all sections across all students who consented to participate in the study (Q4 is best). This comparison is descriptive only; as discussed in Section 3.3, known but unreportable section-level confounding renders the between-group grade comparison uninformative about the demonstrations. Note that unlike Fig. \ref{fig:performance}, this figure is not restricted to students who completed both the pre and post surveys.}   
 \label{fig:demo_imply}
\end{figure}

For completeness, Figures~\ref{fig:demo_imply} reports AI use frequency and grade quartile composition by section group. Reported use was broadly similar across groups: mechanics-specific never-use was close to 33\% in both groups at pre, and both groups shifted strongly toward regular use by post, with no remaining never-users among demo-section respondents (0 of 61) versus 6 of 38 in the other sections. We caution against reading even this modest difference as a demonstration effect. The group samples are small, the results are preliminary, and the outcome is self-reported. For example, students in the demo
section, where AI use was discussed openly all semester, may simply have been more comfortable disclosing their use. The grade comparison warrants stronger caution still. Beyond the composition difference visible in Figure~\ref{fig:demo_imply} (the non-demo group contains a larger share of top-quartile students), the instructor is aware of section-level context that cannot be reported without compromising de-identification, including substantial differences in section enrollment, rates of section-level consent to participate in the study, and in section-level exam performance that do not align with demonstration exposure. Given this unreportable but known confounding, we treat the between-group grade comparison as uninformative about the demonstrations and include it only for completeness and transparency as it is an obvious result to examine in this context.

\section{Conclusion}

This study provides a preliminary empirical snapshot of how undergraduate engineering mechanics students are integrating generative AI into their academic workflows. The purpose of publishing these preliminary results is twofold. First, this study provides a timely and comprehensive dataset specifically focused on generative AI use within the domain of engineering mechanics. Although the modest sample size limits our ability to make causal claims, these descriptive results offer immediate, practical insights for engineering educators. Furthermore, while the specific AI demonstrations deployed in this study will undoubtedly evolve alongside the rapidly advancing underlying technology, the current sequence provides a viable, structured educational resource that other instructors can adapt and implement in their own classrooms today. 

Second, undergraduate education is currently undergoing a rapid, unscripted transformation. We are living through an unprecedented educational experiment, the long-term cognitive and pedagogical consequences of which remain unknown. By making our study design, survey instrument, and demonstration materials publicly available, we hope to establish a reproducible foundation for future research. As our collective understanding of how LLMs influence learning matures, we can begin to ask more sophisticated and targeted questions about how specific instructor initiated guidance around LLM use should be implemented in domain specific educational contexts. 
For example, an early review of pedagogically constrained LLM use finds mostly positive effects on performance, motivation, and higher order thinking, yet observes that most studies compare guided use against no use rather than against free use, leaving the specific pedagogical contribution of guidance design largely unisolated \citep{geng2026}.
We strongly encourage other educators to collect and share similar empirical data within their own specific domains as navigating this shifting landscape will require a collaborative, evidence-based approach.
Our intention is to evolve and repeat this study over multiple semesters while continuing to share the outcomes. 
The harder and more important task ahead is to develop ethical, longitudinal methods for assessing how ubiquitous LLM use and LLM-centered pedagogy reshape engineering education, and we intend for this work to be a first step toward that goal.

\FloatBarrier

\section{Data Availability Statement}
\label{sec:data_availability}

All data required to reproduce the results shown in this manuscript and described in Section \ref{sec:measures} has been published in the OpenBU Institutional Repository at \url{https://hdl.handle.net/2144/53355} \citep{geng_openbu_dataset}. All code required to re-generate the figures in this manuscript and produce the qualitative coding results has been published on GitHub at \url{https://github.com/elejeune11/LLMs_Education_Spring_2026_Data_Analysis}. 

\section{Author Contribution Statement}

Author contributions are specified in Table \ref{tab:credit_roles}. 

\renewcommand{\arraystretch}{1.2} 
\begin{longtable}{l | c c c c c c c}
\caption{Author contributions according to the CRediT (Contributor Roles Taxonomy).}
\label{tab:credit_roles} \\

\hline
\textbf{CRediT Role} & \textbf{SG} & \textbf{HLH} & \textbf{JA} & \textbf{TJM} & \textbf{AD} & \textbf{CF} & \textbf{EL} \\
\hline
\endfirsthead

\multicolumn{8}{c}%
{{\tablename\ \thetable{} -- continued from previous page}} \\
\hline
\textbf{CRediT Role} & \textbf{SG} & \textbf{HLH} & \textbf{JA} & \textbf{TJM} & \textbf{AD} & \textbf{CF} & \textbf{EL} \\
\hline
\endhead

\hline \multicolumn{8}{r}{\textit{Continued on next page}} \\ 
\endfoot

\hline
\endlastfoot

Conceptualization             &X& & & & & &X \\
Data curation                    &X&X&X& & & &X \\
Formal analysis                & & & & & & &X \\
Funding acquisition          & & & & & & &X \\
Investigation                     &X& & & & & &X \\
Methodology                     &X&X&X&X&X&X&X \\
Project administration       &X& & & &X& &X \\
Resources                         & & & & & & &X \\
Software                            & & & & & & &X \\
Supervision                       &X& & & &X& &X \\
Validation                           & & & & & & &X \\
Visualization                      & & & & & & &X \\
Writing -- original draft       & & & & & & &X \\
Writing -- review \& editing &X&X&X&X&X&X&X \\

\end{longtable}

\section{Acknowledgements}
This study was made possible by support from the Boston University Department of Mechanical Engineering, a Shipley Academic Innovation Award from the Institute for Excellence in Teaching \& Learning, and NSF grants CMMI-2311640 and CSSI-2310771. This support is gratefully acknowledged. 

\appendix

\section{Survey Instrument, Spring 2026}
\label{apx:survey}

This Appendix lists every item from the pre- and post-semester Qualtrics surveys, in administration order. PRE items were collected at the start of the semester; POST items were collected at the end of the semester. Items asked at both timepoints used identical wording unless explicitly noted.
 
\textbf{Format.} Each entry is headed by the dataset variable name(s) used in the deposited file. PRE/POST variants are listed when applicable. Item text is shown exactly as presented to participants, followed by response options in the order presented and the response type.
 
\textbf{Mapping to deposited dataset.} Variable names correspond to columns in the de-identified dataset deposited via OpenBU \url{https://hdl.handle.net/2144/53355}. Variables ending in \texttt{\_pre} come from the start-of-semester survey; variables ending in \texttt{\_post} come from the end-of-semester survey.

\subsection*{Preamble}

This survey asks about your experiences and attitudes related to learning in EK301 and the use of AI tools such as TerrierGPT, ChatGPT, Claude, Gemini, DeepSeek etc. Please answer each question honestly and thoughtfully based on your own experience. There are no right or wrong answers.

Your responses are confidential to the instructors and will not negatively affect your grade or standing in the course.

In the questions that follow, “Generative AI tools” refers to tools such as TerrierGPT, ChatGPT, Claude, Gemini, DeepSeek, or similar.

The survey should take approximately 10 minutes to complete.

\subsection*{A. Self-rated understanding of engineering mechanics}
Asked at both PRE and POST. Same wording at both timepoints. 5-point Likert.
 
\noindent\textbf{Item 1:} \verb|understand_apply_principles_pre| (PRE), \verb|understand_apply_principles_post| (POST)\\
\textit{Stem:} Please indicate how much you agree with each statement.\\
\textit{Item:} I feel confident applying engineering mechanics principles to solve new problems\\
\textit{Response options:} Strongly Disagree \textbar{} Disagree \textbar{} Neither Agree nor Disagree \textbar{} Agree \textbar{} Strongly Agree\\
\textit{Type:} 5-point Likert
 
\vspace{0.5em}
 
\noindent\textbf{Item 2:} \verb|understand_identify_mistakes_pre| (PRE), \verb|understand_identify_mistakes_post| (POST)\\
\textit{Stem:} Please indicate how much you agree with each statement.\\
\textit{Item:} I can usually identify and correct my own mistakes when solving engineering mechanics problems.\\
\textit{Response options:} Strongly Disagree \textbar{} Disagree \textbar{} Neither Agree nor Disagree \textbar{} Agree \textbar{} Strongly Agree\\
\textit{Type:} 5-point Likert
 
\vspace{0.5em}
 
\noindent\textbf{Item 3:} \verb|understand_until_practice_pre| (PRE), \verb|understand_until_practice_post| (POST)\\
\textit{Stem:} Please indicate how much you agree with each statement.\\
\textit{Item:} I think that I understand concepts until I start working on specific examples on my own.\\
\textit{Response options:} Strongly Disagree \textbar{} Disagree \textbar{} Neither Agree nor Disagree \textbar{} Agree \textbar{} Strongly Agree\\
\textit{Type:} 5-point Likert
 
\vspace{0.5em}
 
\noindent\textbf{Item 4:} \verb|understand_exams_reflect_pre| (PRE), \verb|understand_exams_reflect_post| (POST)\\
\textit{Stem:} Please indicate how much you agree with each statement.\\
\textit{Item:} My exam scores reflect how well I understand the material in my engineering classes.\\
\textit{Response options:} Strongly Disagree \textbar{} Disagree \textbar{} Neither Agree nor Disagree \textbar{} Agree \textbar{} Strongly Agree\\
\textit{Type:} 5-point Likert
 
\vspace{0.5em}
 
\subsection*{B. AI use frequency}
Asked at both PRE and POST. Same wording at both timepoints. 5-point ordinal.
 
\noindent\textbf{Item 5:} \verb|ai_use_freq_general_pre| (PRE), \verb|ai_use_freq_general_post| (POST)\\
\textit{Stem:} Please indicate your generative AI tool use frequency.\\
\textit{Item:} On average, how often do you use generative AI tools for one or a variety of purposes?\\
\textit{Response options:} Never used \textbar{} Less than once per month \textbar{} 1--3 times per month \textbar{} 1--5 times per week \textbar{} Daily or almost daily\\
\textit{Type:} 5-point ordinal
 
\vspace{0.5em}
 
\noindent\textbf{Item 6:} \verb|ai_use_freq_engmech_pre| (PRE), \verb|ai_use_freq_engmech_post| (POST)\\
\textit{Stem:} Please indicate your generative AI tool use frequency.\\
\textit{Item:} To your best knowledge, how often have you used generative AI tools to help with engineering mechanics coursework specifically?\\
\textit{Response options:} Never used \textbar{} Less than once per month \textbar{} 1--3 times per month \textbar{} 1--5 times per week \textbar{} Daily or almost daily\\
\textit{Type:} 5-point ordinal
 
\vspace{0.5em}
 
\subsection*{C. AI attitudes --- usefulness, confusion, evaluation, accuracy}
Asked at both PRE and POST. Same wording at both timepoints. 5-point Likert with `Not Applicable' option.
 
\noindent\textbf{Item 7:} \verb|ai_useful_coursework_pre| (PRE), \verb|ai_useful_coursework_post| (POST)\\
\textit{Stem:} Please indicate how much you agree with each statement. Select "Not Applicable" if the statement does not apply to you.\\
\textit{Item:} Generative AI tools have been a useful resource for learning university coursework.\\
\textit{Response options:} Strongly Disagree \textbar{} Disagree \textbar{} Neither Agree nor Disagree \textbar{} Agree \textbar{} Strongly Agree \textbar{} Not Applicable\\
\textit{Type:} 5-point Likert with N/A
 
\vspace{0.5em}
 
\noindent\textbf{Item 8:} \verb|ai_helped_understand_engmech_pre| (PRE), \verb|ai_helped_understand_engmech_post| (POST)\\
\textit{Stem:} Please indicate how much you agree with each statement. Select "Not Applicable" if the statement does not apply to you.\\
\textit{Item:} Using AI tools has helped me better understand engineering mechanics.\\
\textit{Response options:} Strongly Disagree \textbar{} Disagree \textbar{} Neither Agree nor Disagree \textbar{} Agree \textbar{} Strongly Agree \textbar{} Not Applicable\\
\textit{Type:} 5-point Likert with N/A
 
\vspace{0.5em}
 
\noindent\textbf{Item 9:} \verb|ai_more_confusing_pre| (PRE), \verb|ai_more_confusing_post| (POST)\\
\textit{Stem:} Please indicate how much you agree with each statement. Select "Not Applicable" if the statement does not apply to you.\\
\textit{Item:} Using AI tools makes engineering mechanics problems more confusing rather than clearer.\\
\textit{Response options:} Strongly Disagree \textbar{} Disagree \textbar{} Neither Agree nor Disagree \textbar{} Agree \textbar{} Strongly Agree \textbar{} Not Applicable\\
\textit{Type:} 5-point Likert with N/A
 
\vspace{0.5em}
 
\noindent\textbf{Item 10:} \verb|ai_can_evaluate_correctness_pre| (PRE), \verb|ai_can_evaluate_correctness_post| (POST)\\
\textit{Stem:} Please indicate how much you agree with each statement. Select "Not Applicable" if the statement does not apply to you.\\
\textit{Item:} I can evaluate whether an AI-generated solution is correct for an engineering mechanics problem.\\
\textit{Response options:} Strongly Disagree \textbar{} Disagree \textbar{} Neither Agree nor Disagree \textbar{} Agree \textbar{} Strongly Agree \textbar{} Not Applicable\\
\textit{Type:} 5-point Likert with N/A
 
\vspace{0.5em}
 
\noindent\textbf{Item 11:} \verb|ai_accuracy_concerns_pre| (PRE), \verb|ai_accuracy_concerns_post| (POST)\\
\textit{Stem:} Please indicate how much you agree with each statement. Select "Not Applicable" if the statement does not apply to you.\\
\textit{Item:} I have concerns about the accuracy of AI tools for solving engineering mechanics problems.\\
\textit{Response options:} Strongly Disagree \textbar{} Disagree \textbar{} Neither Agree nor Disagree \textbar{} Agree \textbar{} Strongly Agree \textbar{} Not Applicable\\
\textit{Type:} 5-point Likert with N/A
 
\vspace{0.5em}
 
\subsection*{D. AI relative to other learning resources}
Asked at both PRE and POST. Same wording at both timepoints. 5-point Likert with `Not Applicable' option.
 
\noindent\textbf{Item 12:} \verb|ai_vs_textbooks_pre| (PRE), \verb|ai_vs_textbooks_post| (POST)\\
\textit{Stem:} Please indicate how much you agree with each statement. Select "Not Applicable" if the statement does not apply to you.\\
\textit{Item:} Overall, AI tools are more helpful than textbooks for learning engineering mechanics.\\
\textit{Response options:} Strongly Disagree \textbar{} Disagree \textbar{} Neither Agree nor Disagree \textbar{} Agree \textbar{} Strongly Agree \textbar{} Not Applicable\\
\textit{Type:} 5-point Likert with N/A
 
\vspace{0.5em}
 
\noindent\textbf{Item 13:} \verb|ai_vs_discussion_pre| (PRE), \verb|ai_vs_discussion_post| (POST)\\
\textit{Stem:} Please indicate how much you agree with each statement. Select "Not Applicable" if the statement does not apply to you.\\
\textit{Item:} Overall, AI tools are more helpful than the course's discussion section (also referred to as weekly office hours) for learning engineering mechanics.\\
\textit{Response options:} Strongly Disagree \textbar{} Disagree \textbar{} Neither Agree nor Disagree \textbar{} Agree \textbar{} Strongly Agree \textbar{} Not Applicable\\
\textit{Type:} 5-point Likert with N/A
 
\vspace{0.5em}
 
\subsection*{E. AI impact rating}
Asked at both PRE and POST. Same wording at both timepoints.
 
\noindent\textbf{Item 14:} \verb|ai_impact_rating_pre| (PRE), \verb|ai_impact_rating_post| (POST)\\
\textit{Item:} How would you rate the impact of AI tools on your learning in engineering mechanics?\\
\textit{Response options:} 0 (very negative) to 10 (very positive)\\
\textit{Type:} 0--10 radio buttons
 
\vspace{0.5em}
 
\subsection*{F. AI demonstration experience items (POST only)}
Asked at POST only. \verb|ai_demos_attended_post| was asked of all students; the four items that follow were branched only to students who reported attending one or more demonstrations.

\noindent\textbf{Item 15:} \verb|ai_demos_attended_post|\\
\textit{Stem:} To your best knowledge, how many of the 9 AI demonstrations did you attend? If you were in a section that did not have AI demonstrations, select 0.\\
\textit{Item:} Estimate\\
\textit{Response options:} Integer 0--9\\
\textit{Type:} 0--9 radio buttons
 
\vspace{0.5em}
 
\noindent\textbf{Item 16:} \verb|ai_demo_helped_understanding_post|\\
\textit{Item:} The AI demonstrations in class helped strengthen my understanding of how to solve engineering mechanics problems.\\
\textit{Response options:} Strongly Disagree \textbar{} Disagree \textbar{} Neither Agree nor Disagree \textbar{} Agree \textbar{} Strongly Agree\\
\textit{Type:} 5-point Likert 
 
\vspace{0.5em}
 
\noindent\textbf{Item 17:} \verb|ai_demo_engagement_post|\\
\textit{Item:} On average, how engaged were you when attending AI demonstrations?\\
\textit{Response options:} Not at all engaged \textbar{} Slightly engaged \textbar{} Moderately engaged \textbar{} Very engaged \textbar{} Extremely engaged\\
\textit{Type:} 5-point ordinal
 
\vspace{0.5em}
 
\noindent\textbf{Item 18:} \verb|ai_demo_changed_use_post|\\
\textit{Item:} To what extent did the AI demonstrations change how you use AI tools?\\
\textit{Response options:} Not at all \textbar{} A little \textbar{} Somewhat \textbar{} Quite a bit \textbar{} A great deal\\
\textit{Type:} 5-point ordinal
 
\vspace{0.5em}
 
\noindent\textbf{Item 19:} \verb|txt_ai_demo_feedback_post|\\
\textit{Item:} What did you like/dislike about the AI demonstrations you attended? If you have one suggestion to help us improve the AI demonstrations, what would it be?\\
\textit{Response options:} Free-text, open-ended\\
\textit{Type:} Free-text
 
\vspace{0.5em}
 
\subsection*{G. Free-text reflections}
\verb|txt_ai_workflow_steps|, \verb|txt_ai_verify_correctness|, and \verb|txt_ai_impact_describe| asked at both PRE and POST with identical wording. \verb|txt_ai_use_change_post| asked at POST only. No character limit imposed.
 
\noindent\textbf{Item 20:} \verb|txt_ai_workflow_steps_pre| (PRE), \verb|txt_ai_workflow_steps_post| (POST)\\
\textit{Item:} When you use AI tools for this course, what steps do you usually take from first question to final answer?\\
\textit{Response options:} Free-text, open-ended\\
\textit{Type:} Free-text
 
\vspace{0.5em}
 
\noindent\textbf{Item 21:} \verb|txt_ai_verify_correctness_pre| (PRE), \verb|txt_ai_verify_correctness_post| (POST)\\
\textit{Item:} What, if anything, do you do to check whether an AI-generated response is correct?\\
\textit{Response options:} Free-text, open-ended\\
\textit{Type:} Free-text
 
\vspace{0.5em}
 
\noindent\textbf{Item 22:} \verb|txt_ai_impact_describe_pre| (PRE), \verb|txt_ai_impact_describe_post| (POST)\\
\textit{Item:} How would you describe the impact of AI tools (positive and negative) on your learning in this course?\\
\textit{Response options:} Free-text, open-ended\\
\textit{Type:} Free-text
 
\vspace{0.5em}
 
\noindent\textbf{Item 23:} \verb|txt_ai_use_change_post| (POST)\\
\textit{Item:} Please describe any changes in how you use AI tools for your learning since the start of the semester.\\
\textit{Response options:} Free-text, open-ended\\
\textit{Type:} Free-text
 
\vspace{0.5em}
 
\subsection*{H. Response validity}
Asked at both PRE and POST. Same wording at both timepoints. 5-point Likert.
 
\noindent\textbf{Item 24:} \verb|responded_honestly_pre| (PRE), \verb|responded_honestly_post| (POST)\\
\textit{Stem:} Please indicate how much you agree with this statement.\\
\textit{Item:} I responded honestly and thoughtfully to each question in this survey.\\
\textit{Response options:} Strongly Disagree \textbar{} Disagree \textbar{} Neither Agree nor Disagree \textbar{} Agree \textbar{} Strongly Agree\\
\textit{Type:} 5-point Likert
 
\vspace{0.5em}
 
\subsection*{Notes on removed question from PRE survey}

Asked at PRE only. Item framed around prior exposure to AI demonstrations; students could select Not Applicable if not previously exposed. This question was confusing, and was administered before any AI demonstrations occurred, so we removed it from our dataset. 
 
\noindent \textbf{Removed item:} \verb|ai_demo_helped_understanding_pre| (PRE)\\
\textit{Stem:} Please indicate how much you agree with this statement. Select "Not Applicable" if you were not exposed to the AI Demonstrations.\\
\textit{Item:} The AI demonstrations in class helped strengthen my understanding of how to solve engineering mechanics problems.\\
\textit{Response options:} Strongly Disagree \textbar{} Disagree \textbar{} Neither Agree nor Disagree \textbar{} Agree \textbar{} Strongly Agree \textbar{} Not Applicable\\
\textit{Type:} 5-point Likert with N/A

\section{Details of AI Demonstrations, Spring 2026}
\label{apx:demos}

This Appendix documents the nine AI demonstrations delivered in the intervention section during Spring 2026. For each demonstration, we provide the pedagogical purpose, a description of the planned content, a summary of live LLM interactions, and instructor reflections. All live LLM interactions used Gemini in the web browser, with prior conversation histories cleared before each demonstration to ensure a completely fresh model context. These instructor reflections are a consolidation of notes taken throughout the semester prior to the instructor seeing any of the survey results. We structure these written descriptions to convey the pedagogical intent behind each demonstration in a manner that would be transferable to other educators. In the next semester of this research study, we plan to significantly revise the demonstrations based on student feedback, instructor experience, and the changing landscape of LLM abilities. 

\subsection*{Demonstration 1: LLM Use in Group Projects} 

\textbf{Pedagogical purpose:} To demonstrate instructor familiarity with current LLM capabilities, and to encourage students to discuss whether to include ground rules around LLM use in their required group contracts for the semester-long truss design project. The framing was oriented toward peer-to-peer responsibility rather than academic integrity policy. Specifically, what do you owe your groupmates when you use an LLM?

\textbf{Description:} This demonstration was delivered during the introductory lecture for the truss design project, which begins with a ``buckling lab'' in which students experimentally measure the critical buckling load of acrylic strips as a function of length to characterize the material for subsequent truss design. The demonstration began with a brief discussion of model selection, referencing FEM-Bench performance data \citep{mohammadzadeh2025fem} to motivate why a particular model (Gemini, accessed directly from Google) was chosen over other options available through the university-provided LLM interface (TerrierGPT). The instructor then uploaded the buckling lab project manual and a required case study document on group dynamics to Gemini and ran three prompts live. The demonstration concluded with a five-minute small-group discussion: should ground rules around LLM use be a core component of the group contract?

\textbf{Live LLM interactivity:} Three prompts were run in Google Gemini (Gemini 3 Pro, Fast mode) with the project manual and case study uploaded as context. The first prompt, ``write a draft lab report based on the lab manual and associated case study document,'' produced a structured report closely following the rubric, with appropriate section headings and placeholder values for experimental data. The second prompt, which invoked Nano Banana (Gemini's image generation model), ``generate the free body diagram required for the lab report,'' produced an image of a free body diagram of the loaded acrylic strip on a tared scale (notably, this was not a correct free body diagram). The third prompt, which invoked Nano Banana, ``generate the plot required for the lab report,'' produced a plot of critical buckling load versus strip length with fabricated data, a linear fit, and visible errors in the axis labels.

\textbf{Instructor reflection:} In the instructor's read of the room, students were entirely unsurprised that an LLM could produce a near-complete lab report. In the group discussion, students appeared to reach quick consensus that LLM use on the project was acceptable but should be disclosed to groupmates. No students advocated banning LLMs. The instructor raised the concern that LLMs might interfere with learning how to write a lab report; in the instructor's read students did not appear particularly concerned about this. The demonstration did appear to make students more comfortable discussing their own AI use openly, which the instructor hoped would yield more candid engagement in subsequent demonstrations.

\subsection*{Demonstration 2: LLMs + The Python Sandbox}

\textbf{Pedagogical purpose:} To introduce students to the concept of a Python sandbox (an isolated code execution environment embedded within an LLM interface), to highlight differences in capability across LLM providers, and to help students think critically about when an LLM interface's built-in code execution may be useful to them.

\textbf{Description:} The demonstration began by defining what a Python sandbox is and why LLMs have Python sandboxes but not MATLAB sandboxes (Python is open-source; MATLAB is proprietary). The instructor then showed the same prompt, ``Do you have a python sandbox?'', submitted to multiple models available through BU's TerrierGPT platform (OpenAI GPT-5.1, Anthropic Claude Sonnet 4.5, and Google Gemini 3 Flash) as well as to the standalone Gemini 3 Flash interface. The TerrierGPT-hosted models all reported lacking a sandbox, while standalone Gemini confirmed sandbox access. This comparison served as a concrete illustration that LLM interactions will vary both by model and by model interface. The instructor then demonstrated Gemini's sandbox by uploading a synthetic dataset and iteratively prompting it to generate and refine plots. The demonstration concluded with a discussion of the pros (convenience, live debugging, safety of sandboxed execution) and cons (limited persistence, cost, potential tedium of iterative prompting for small changes) of using an LLM's Python sandbox versus running code locally.

\textbf{Live LLM interactivity:} Prompts were run in Google Gemini 3 Flash with a synthetic CSV dataset uploaded as context. The first prompt, ``can you write python code to make a scatter plot of these data then run the code?'', produced a scatter plot with the generated Python code displayed alongside it. The second prompt, ``can you turn this into a box and whiskers plot for each of the four values of x?'', produced an updated visualization using seaborn. The instructor had planned additional iterative refinements (e.g., relabeling axes with engineering units, overlaying raw data) but curtailed the sequence in class when it became clear that the iterative prompting pattern was self-evident to students.

\textbf{Instructor reflection:} In the instructor's assessment, this was likely the most successful of the early demonstrations. An informal poll indicated that most students had not previously encountered the concept of a Python sandbox, suggesting genuine new learning. Students appeared engaged and interested in the practical skill being demonstrated. In retrospect, the demonstration could have been more explicitly framed around the concept of ``tool use'' (i.e., LLMs invoking external tools).

\subsection*{Demonstration 3: The ``Zone of Proximal Development''}

\textbf{Pedagogical purpose:} To introduce students to the concepts of metacognition and the Zone of Proximal Development (ZPD) \citep{Vygotsky1978}, and to facilitate a structured discussion about whether LLMs can effectively serve as a ``More Knowledgeable Other'' in the context of learning engineering mechanics.

\textbf{Description:} This demonstration was delivered as a conceptual discussion rather than a live LLM interaction. The instructor introduced metacognition (``thinking about thinking'') and the ZPD, defined as the space between what a learner can do independently and what they cannot do even with support \citep{Vygotsky1978}. The instructor connected these concepts to EK301 specifically, noting that group problems, homework, and the truss project are all designed to place students in their ZPD, and that the typical ``More Knowledgeable Others'' in EK301 are the professor, graduate student teachers, undergraduate learning assistants, and peers. The demonstration then posed the question: can LLMs serve as a More Knowledgeable Other? Students discussed in small groups for approximately five minutes, then the class compiled pros (e.g., availability, non-judgmental, large knowledge base) and cons (e.g., hallucinations, bypassing productive struggle, diminishing social learning). The demonstration closed by previewing the next session, which would examine published research on LLMs and learning.

\textbf{Live LLM interactivity:} None. This demonstration was entirely discussion-based.

\textbf{Instructor reflection:} In the instructor's assessment, students were largely unfamiliar with metacognition and the ZPD as concepts but engaged with them quickly. Two contributions during the group discussion stood out. First, a student identified cost as a major pro of LLMs relative to traditional instruction (e.g., the cost of a ChatGPT subscription versus the cost of a university education); the instructor acknowledged this and noted the precedent of MOOCs, which promised similar access and saw massive attrition \citep{onah2014dropout}. Second, students pushed back on the concept of ``productive struggle'' in two ways: one noted scenarios where LLMs saved them from struggle that was annoying rather than productive (e.g., tracking down information), and another argued that LLMs are changing the shape of the productive struggle itself, shifting it toward debugging and verifying LLM output. The instructor responded that whether this shifted form of struggle constitutes effective pedagogy is an open question.

\subsection*{Demonstration 4: LLMs for Learning + System Prompts}

\textbf{Pedagogical purpose:} To confront the reality of how generative AI tools can impede learning when used without guardrails, to present empirical evidence from the learning sciences literature to higher education students, and to demonstrate how custom system prompts can reshape an LLM from a standard ``knowledge oracle'' into an interactive Socratic tutor designed to preserve the user's ``productive struggle.''

\textbf{Description:} Building directly on the theoretical framework established in Demonstration 3, this session transitioned from a conceptual discussion of the Zone of Proximal Development (ZPD) to empirical literature and practical tool modification \citep{Vygotsky1978}. The instructor presented a curated digest of three recent publications evaluating LLM efficacy in education: an overarching literature review \citep{shi2025large}, a large-scale randomized controlled trial revealing that unguarded LLM access significantly degraded subsequent independent exam performance \citep{bastani2025generative}, and recent institutional work on fine-tuning foundational models specifically for educational deployment \citep{team2025ai}. The instructor shared a candid evaluation of these platforms, noting the gap between commercial marketing and classroom utility. To show how users can take agency over their own cognitive guardrails, the instructor introduced the mechanics of system prompts; based on an informal classroom poll, this concept was unfamiliar to approximately 80\% of the class. The instructor demonstrated a custom-authored system prompt (see Appendix \ref{apx:prompt}) designed to bind Google Gemini into a disciplined, multi-phase EK301 Socratic Tutor implemented as a Gemini GEM. The session concluded by contrasting how a standard LLM deployment immediately bypasses learning by supplying numerical solutions against how a system-prompted model enforces structural statics workflows.

\textbf{Live LLM interactivity:} Both interactions were executed in Google Gemini (Gemini 3 Pro) using an example engineering mechanics problem. In the baseline interaction, the statics problem statement and accompanying diagram were submitted to the default model interface. The default LLM immediately generated a complete numerical solution, outputting algebra and final values without any friction. In the enhanced interaction, the instructor initialized a dedicated model session embedded with a comprehensive system prompt. This prompt explicitly instructed the model to prioritize disciplined problem-solving habits over answers, forcing it to navigate a strict four-phase sequence: high-level planning, system isolation/FBD discipline, symbolic equilibrium equations, and physical plausibility checking. When the identical statics problem was uploaded, the system-prompted model refused to supply answers. Instead, it initiated Phase 1, asking a single, high-level question: \textit{``Before computing anything, what are the major steps you will need to complete to solve this problem?''}. The instructor demonstrated live student-side input to illustrate how the model maintains its guardrails, offering marginal scaffolding hints while prompting the user to independently identify support forces and define coordinate systems before writing equilibrium expressions.

\textbf{Instructor reflection:} Introducing empirical research papers to an undergraduate statics service course was a stylistic gamble, and it resonated well with some students but not others. In the instructor's read, some students appreciated the distinction between immediate workplace productivity gains and long-term cognitive skill acquisition. Because the vast majority of the class had never encountered system prompting, this lecture offered an immediate, actionable takeaway. The live contrast between the standard LLM response and the Socratic prompt was highly apparent. Following the demonstration, multiple students requested the raw text of the system prompt to deploy in their own study workflows. The instructor distributed the prompt via email with a clear caveat: it remains an unvalidated, ad-hoc intervention rather than a rigorously proven educational tool. Based on preliminary investigation, from the instructor's point of view it seems that an off-the-shelf foundational model paired with a meticulously designed system prompt can likely compete directly with highly engineered, proprietary tutor chatbots.

\subsection*{Demonstration 5: Strategic Delegation to LLMs}

\textbf{Pedagogical purpose:} To model how students can strategically delegate low-risk programming tasks (such as graphical user interface development) to LLMs to build high-leverage engineering tools, and to encourage students to go beyond the baseline requirements of a design project by creating an interactive, visual workflow.

\textbf{Description:} This demonstration was delivered as students began the semester-long truss design project. The current infrastructure of the project requires students to manually translate a truss geometry into a system of matrix equations to solve for member forces. The instructor introduced the concept of ``strategic delegation'', i.e., identifying low-risk aspects of a programming workflow and offloading them entirely to an LLM, while keeping the core engineering logic closely managed by the human designer (i.e., in this case the actual truss solver). The instructor showcased iterative prompting to implement a web-based frontend using Streamlit. The instructor demonstrated how the app layout offered interactive data editors for live nodal and member connectivity input, displayed real-time material limits, rendered supports graphically via Matplotlib, and exposed a download trigger for the resulting truss design.

\textbf{Live LLM interactivity:} The instructor input the prompt below to Gemini 3 Pro to model an initial tool-building sequence: 

\begin{quote}
\small\raggedright\texttt{Here is a project manual. I want to create two python files. One is titled ``truss\_project\_code.py'' and runs all of the analysis specified in the truss project manual and saves output accordingly. The other is titled ``truss\_project\_gui.py'' and I want this to be a GUI created with streamlit that allows the user to enter node information (node number, x coordinate, y coordinate, pin/roller/no support condition), element connectivity information, nodal loading information, and then hit run to make the analysis happen. Entering information should be made easy, maybe a table format? something that allows easy changes and updates during the process. The GUI should plot the truss as it is being created, and then display the results of the analysis on the screen in a meaningful way. The GUI should also start with an example truss set up that is easy to modify.}
\end{quote}

This prompt was structured to explicitly emphasize a clean separation of concerns. The instructor highlighted to the class that students retain absolute responsibility for the engineering physics backend (\texttt{truss\_project\_code.py}), as a core learning objective of the project assignment is the independent implementation and rigorous verification of their own underlying analysis code. By contrast, the live interaction demonstrated how students can shift focus toward using LLMs to rapidly build an intuitive layout and a functional design interface (\texttt{truss\_project\_gui.py}) around their verified solver. Following this prompt sequence, the generated code files were downloaded directly from the model interface. The operational Streamlit app was then launched locally inside VS Code using a virtual environment manager to demonstrate the completed interface in front of the class.

\textbf{Instructor reflection:} The timing of this demonstration was somewhat misaligned with student readiness. Because it was delivered before the students fully understood the fundamental mathematical constraints and mechanics of the truss project itself, the core utility of a GUI for truss design seemed that it might have been lost on a significant portion of the cohort. Furthermore, due to time constraints, the pace of the lecture was rushed, preventing the instructor from pausing to gauge student familiarity with the development stack being displayed (such as VS Code, virtual environments, or Streamlit). In the instructor's assessment, a more effective deployment would give students more time to get oriented with the project first, followed by explicit check-ins regarding which programming tools and concepts are familiar or unfamiliar to them. 

\subsection*{Demonstration 6: Strategic Delegation to LLMs Part II}

\textbf{Pedagogical purpose:} To demonstrate how advanced algorithmic techniques (such as optimization routines) can be implemented via LLMs to enhance engineering design iteration, and to inspire students to look beyond baseline design requirements by showcasing how to use automated workflows to explore a complex design space.

\textbf{Description:} This demonstration built directly upon the interactive web-frontend framework established in Demonstration 5. While the previous session focused on strategic delegation of low-risk user interface layout tasks, this session advanced to deploying an algorithmic optimization tool within the design workflow. The instructor introduced the concept of automated iteration, contrasting traditional manual ``trial-and-error'' truss design process against programmatic optimization. Due to the complexity and computational iteration required to build a functional optimization loop, the instructor did not generate the code live in class. Instead, the instructor showcased a pre-developed Streamlit application built through prior iterative prompting. The application combined the baseline statics solver with a simple Monte Carlo optimization routine designed to randomly perturb internal node coordinates within user-defined geometric boundaries. The instructor demonstrated the live optimization process to the class, showing how the Monte Carlo simulation technique could be used to improve the load-to-cost ratio of a baseline truss design.

\textbf{Live LLM interactivity:} Because generating a robust optimization loop requires an extensive, multi-step debugging dialog that is impractical during a live lecture, the instructor instead presented the core high-level prompting strategy used to generate the script before class via slides of a successful prompting sequence. The operational interface (\texttt{truss\_project\_optimize\_gui.py}) was then executed locally to show the real-time analysis and the final optimized truss configuration emerging live on the screen.

\textbf{Instructor reflection:} Following the pacing difficulties encountered in Demonstration 5, where the class felt slightly disconnected from the material, this session successfully recaptured student engagement. Shifting away from a live-coding format allowed the instructor to slow down the delivery and focus heavily on the conceptual ``why'' behind the tool. Visualizing an optimization algorithm actively working to minimize cost in real time served as a strong hook; students appeared genuinely excited by the prospect of using an LLM to build sophisticated tools that could directly optimize their actual course project designs. 

\subsection*{Demonstration 7: Benchmarking LLMs}

\textbf{Pedagogical purpose:} To introduce students to the foundational concept of benchmarking in the context of large language models, and to demonstrate an ad hoc approach to evaluating model responses by individually uploading core engineering mechanics coursework problems.

\textbf{Description:} This demonstration shifted focus toward analyzing the performance limits and reliability of generative AI tools. The instructor introduced the concept of benchmarking, highlighting popular examples (e.g., news headlines about LLMs passing famous exams) and standard evaluation methods documented in current literature. To demonstrate model evaluation in practice, the instructor established an ad hoc benchmark tailored specifically to the EK301 curriculum. This process involved selecting individual course problems, uploading them to an LLM, and directly inspecting the text and visual responses against verified solutions to evaluate structural correctness.

\textbf{Live LLM interactivity:} Interactions were performed by individually uploading specific statics problems into the LLM interface. The model was prompted to process the problem descriptions and accompanying figures to generate solutions. The live demonstration focused on reviewing these responses in real time, enabling the class to qualitatively analyze the accuracy of the model's structural logic and final algebraic outcomes against the established course solutions.

\textbf{Instructor reflection:} The goal of this demonstration was to communicate the key ideas behind model evaluation to lay the groundwork for a more technical, programmatic benchmarking session in the subsequent lecture. The instructor was sincerely unsure if the premise of ``quizzing LLMs'' or comparing LLM responses to known solution keys resonated with students. 

\subsection*{Demonstration 8: Benchmarking LLMs Part II}

\textbf{Pedagogical purpose:} To demonstrate the foundational mechanics of building a structured, programmatic scientific benchmark for large language models, drawing from open-source research code to show how automation, multiple model runs, and API calls are used to robustly evaluate course content. Furthermore, establishing automated benchmarks is an increasingly relevant workforce skill, as we anticipate that modern engineering employers will require rigorous verification methodologies to determine whether autonomous workflows can be safely integrated into professional practice. 

\textbf{Description:} This demonstration expanded on the foundational concepts introduced in Demonstration 7 by transitioning from manual, qualitative ad-hoc testing to programmatic evaluation. The instructor introduced how a structured scientific benchmark is constructed from the ground up to evaluate large language models on engineering course content. The instructor quickly walked through the architecture of a custom benchmarking system designed for EK301 course material, explaining how it integrates problem loading, automated API calls to large language models, multiple identical problem iterations to capture statistical variability, and automated answer evaluation to bypass the need for manual inspection. The instructor also noted that in the demonstration benchmark flagship LLMs performed better on the textbook problems included in the benchmark than on the original problems.  

\textbf{Live LLM interactivity:} None. No live code or API calls were executed during the class session. Instead, the demonstration was purely presentation-based, using slides to step through the code structure and show the process of constructing the automation pipeline, establishing model endpoints, and programmatically plotting aggregate model performance results.

\textbf{Instructor reflection:} Overall, despite the instructor's excitement for this topic, only a small number of students felt fully engaged in this concept. One potential source of this challenge might have been student fatigue towards the end of the semester. That being said, the instructor felt that building out a more comprehensive version of this preliminary Engineering Mechanics benchmark could be an impactful teaching resource if presented correctly.

\subsection*{Demonstration 9: Course Wrap-Up and Student Feedback}

\textbf{Pedagogical purpose:} To close the loop on the semester-long intervention by prompting students to reflect critically on the full sequence of AI demonstrations, and to gather aggregate student feedback regarding the perceived utility and impact of these tutorials on their learning workflows.

\textbf{Description:} This final demonstration served as the conclusion to the classroom research study. Rather than presenting new technical workflows or software code, the instructor dedicated the session to an open discussion summarizing all prior demonstrations delivered throughout the semester. The instructor distributed a summary sheet listing the complete tutorial index to help students contextualize the progression from early tool literacy to algorithmic delegation and systematic benchmarking. Students were then asked to participate in an interactive, aggregate survey to rank which specific demonstrations they found most useful, providing qualitative and quantitative data to guide future iterations of the course curriculum.

\textbf{Live LLM interactivity:} None. This session was entirely feedback-driven and reflection-based. The instructor displayed slides summarizing the primary themes of the nine tutorials and facilitated a live polling session to capture student sentiment regarding the overall efficacy of the intervention. 

\textbf{Instructor reflection:} The collection of end-of-semester feedback provided invaluable insights into how different cohorts of students internalized these interventions. In the instructor's read of the aggregate results, student sentiment was highly varied; certain tutorials resonated deeply with some students while leaving others completely disengaged. This disparity underscored the diverse landscape of student backgrounds, prior programming exposure, and individual study habits within a large service course. In retrospect, relying solely on a comprehensive, end-of-term wrap-up survey made it difficult to capture the immediate, nuanced cognitive shifts occurring after each individual intervention. Gathering continuous feedback immediately following each demonstration throughout the semester would have yielded much richer data and allowed for real-time course corrections. These takeaways, combined with the final survey metrics, establish a clear foundation for majorly restructuring the pacing, delivery style, and baseline technical expectations of these AI demonstrations in subsequent semesters of this research study.

\section{Socratic Tutor System Prompt Text}
\label{apx:prompt}

\begin{lstlisting}
Name: Guided Socratic Tutor Engineering Mechanics

Instructions:

You are a guided Socratic tutor for an undergraduate Engineering Mechanics: Statics course.

Your objective is to train disciplined problem-solving habits, not merely obtain correct numerical answers.

Students will upload problems, often with diagrams. You must guide them to independently execute a complete statics workflow.

Core Pedagogical Structure

Every problem must follow this high-level sequence:

Interpret the question.

Identify the system of interest.

Construct a Free Body Diagram (FBD).

Write equilibrium equations.

Solve symbolically.

Check physical plausibility.

You must reinforce this structure through questioning.

Phase 1: High-Level Planning (Before Any Details)

At the beginning of every new problem, ask a single high-level question that prompts the student to articulate the full solution strategy.

Examples of acceptable initial prompts:

``Before computing anything, what are the major steps you will need to complete to solve this problem?''

``How will you structure your approach from interpreting the diagram to obtaining the final answer?''

``What is your plan before writing any equations?''

Do not mention the Free Body Diagram explicitly in this first question.

The goal is for the student to recall it independently.

If they omit a critical step (such as the FBD), ask:

``Are you missing any structural step between identifying the system and writing equations?''

Only introduce the FBD explicitly if they repeatedly fail to identify it.

Phase 2: System Isolation and FBD Discipline

You must never allow equilibrium equations to be written before the system is clearly defined and isolated.

Before permitting equations, ensure the student has:

Clearly stated the body or subsystem being isolated.

Identified all external forces and moments.

Chosen a coordinate system.

If they attempt equations prematurely, respond with a question such as:

``What does your Free Body Diagram look like?''

``Have all external forces acting on your isolated system been identified?''

``What forces appear because of your chosen supports?''

Do not provide the FBD yourself unless the student explicitly requests full explanation after multiple failed attempts.

Phase 3: Equilibrium Equation Discipline

Once the FBD is complete:

Ask which equilibrium equations are appropriate.

Require symbolic equations before substitution.

Ensure correct sign conventions.

If an equation is incorrect:

Ask which force or moment may be missing.

Ask how sign conventions were chosen.

Ask whether the equation reflects the FBD exactly.

Phase 4: Physical Reasoning and Plausibility

After a solution is obtained, always ask a reflective question such as:

``Does this magnitude and direction make physical sense?''

``What would happen if the load increased?''

``What limiting case could you test?''

Statics reasoning must include physical intuition, not just algebra.

Confirmation and Correction Rules

Confirm correctness briefly only after the student has:

Completed all steps independently.

Produced internally consistent equations.

Obtained a physically plausible result.

If the student reaches a stable but incorrect result:

Ask a question exposing the structural flaw.

If necessary after multiple attempts, give a concise correction.

Immediately return to questioning.

Strict Constraints

Ask one primary question per response.

Do not provide full worked solutions unless explicitly requested.

Do not draw diagrams for the student unless they request complete explanation after failed attempts.

Do not skip structural steps even if the algebra is simple.

Maintain intellectual seriousness; avoid praise.
\end{lstlisting}

\section{Codebooks}
\label{apx:codebook}

\subsection{Codebook for Item 19, \texttt{txt\_ai\_demo\_feedback\_post}}

\input{tables/codebook_item_19.tex}

\subsection{Codebook for Item 20, \texttt{txt\_ai\_workflow\_steps\_pre} and \texttt{txt\_ai\_workflow\_steps\_post}}

\input{tables/codebook_item_20.tex}

\subsection{Codebook for Item 21, \texttt{txt\_ai\_verify\_correctness\_pre} and \texttt{txt\_ai\_verify\_correctness\_post}}

\input{tables/codebook_item_21.tex}

\subsection{Codebook for Item 22,  \texttt{txt\_ai\_impact\_describe\_pre} and \texttt{txt\_ai\_impact\_describe\_post}}

\input{tables/codebook_item_22.tex}

\subsection{Codebook for Item 23, \texttt{txt\_ai\_use\_change\_post}}

\input{tables/codebook_item_23.tex}

\section{LLM-Assisted Qualitative Coding Protocol}
\label{apx:llm_coding}

This Appendix documents the pipeline used to code the free-text survey responses (Items 19--23), in sufficient detail to reproduce it. Machine coding of qualitative data has strong potential to democratize this form of research, and we view this pipeline as a way to lower the barriers to studying how students use AI. By publishing both the code and data required to reproduce our results, we aim to support analysis at scale as more data emerges. Because the same pipeline yields codings comparable across courses and institutions, it also creates an impetus to collect and share data and data processing workflows, turning isolated course-level snapshots into a shared evidence base open to the broad community of educators already navigating this shift, not only to well-resourced and siloed research groups. The complete pipeline code, prompts, codebooks, and per-model outputs are included in the public GitHub repository associated with this project. 

\subsection{Codebook development}
Codebooks were drafted by the research team from a full reading of the response corpus and structured uniformly: each code specifies an identifier, a category, apply-when criteria, do-not-apply criteria, distinguish-from notes contrasting it with its most confusable neighbors, and example responses drawn from the data. Based on the emergent trends in the data, two codebooks are tiered, with a mutually exclusive Tier 1 code capturing the response's overall classification and select-all-that-apply Tier 2 codes capturing specifics; the remaining three are flat multi-label codebooks. Codebooks were refined across two pilot
coding rounds (5 and 25 responses per item) before the full run. Pilot revisions were driven by observed inter-model disagreement patterns; in particular, repeated use of residual (``Other'') codes by multiple independent LLM coders was treated as evidence of a missing category, which led to the addition of codes for one-shot answer generation and attempt-first help-seeking workflows (Item 20) and for affective and confidence-related benefits (Item 22). The codebooks were frozen before the full coding run; Appendix D reproduces each code's category and description, and the deposited materials include the complete versions with decision rules and examples. 

\subsection{Coding procedure}
Each (response, coder) pair was one API call. The system prompt contained the survey question as shown to students, the full codebook, the coding rules (including exclusivity and tier-cardinality constraints), and a strict JSON output schema comprising the assigned code identifiers, a one-sentence rationale citing the response's wording, a boolean uncertainty flag, and a notes field required whenever a residual code was applied. The three coding models were \texttt{claude-sonnet-5} (Anthropic), \texttt{gpt-5.5} (OpenAI), and \texttt{gemini-3.5-flash} (Google), accessed via their public APIs in July 2026. Sampling parameters were left at provider defaults because current frontier models reject or discourage non-default values. Model outputs were validated programmatically (known code identifiers only, at least one code, tier cardinality respected, standalone codes uncombined); an invalid output triggered one automated repair attempt with the validation error appended, and persistent failures were logged and routed to adjudication. Coders were blind to one another, to respondent identity, and to section assignment; only the study identifier and response text were transmitted to the API providers, under no-training API terms.

\subsection{Adjudication}
Responses with identical code sets from all three coders and no uncertainty flags were accepted directly (445 of 756 response-question pairs, 58.9\%). All other responses were adjudicated by \texttt{claude-fable-5}, which received the response, the codebook, and each coder's codes and rationales, with instructions to keep, drop, or add codes strictly according to the codebook definitions rather than by majority or deference; adjudicator output passed the same validation rules, and all 311 adjudication calls completed successfully (a majority-vote fallback existed for adjudication failures but was never invoked). Because the adjudicator shares a provider with one of the three coders, we audited for self-preference by computing, over the 311 adjudicated items, the mean Jaccard similarity, defined for two code sets $A$ and $B$ as $J(A,B) = |A \cap B| \, / \, |A \cup B|$, between the final ruling and each coder's original codes: 0.68 for the Anthropic coder, 0.81 for the OpenAI coder, and 0.77 for the Google coder. The adjudicator agreed least, not most, with its same-provider coder, indicating no direct evidence of self-preference in its rulings.

\subsection{Reliability}
Treating each code as a binary apply/not-apply decision per response, we computed Fleiss' $\kappa$, defined as
$\kappa = (\bar{P} - \bar{P}_e)/(1 - \bar{P}_e)$, where $\bar{P}$ is the mean observed pairwise agreement across responses and $\bar{P}_e$ the agreement expected by chance given each coder's marginal code prevalence, across the three coders for every code, restricted to the 755 of 756 response-question pairs coded successfully by all three (for one response, one coder's calls failed output validation persistently; that response retained two valid codings, passed through the standard adjudication path, and was excluded only from the reliability computation). Because $\kappa$ is unstable for rare codes, we report it only for codes with mean prevalence of at least 5\%; prevalence and raw percent agreement are reported for all codes. Table~\ref{tab:llm_reliability} summarizes reliability by question; Table~\ref{tab:llm_reliability_by_code} reports per-code statistics. Across questions, responses receiving identical code sets from all three coders ranged from 38\% (workflow steps, post) to 82\% (change in AI use), tracking coding density: the workflow item averaged 2.6 codes per response against 1.2 for the use-change item, and set identity is broken by any single peripheral disagreement even when the core codes agree. The five codes falling below $\kappa = .61$ all combine prevalence between 5\% and 8\% with raw percent agreement of 0.93 or higher.

\input{tables/table_reliability.tex}
\input{tables/table_reliability_by_code.tex}

\subsection{Scope and validation status}
Machine-coder agreement, however high, establishes reliability rather than validity: three models and an adjudicator can err in the same direction. We therefore treat the coded frequencies in Section \ref{sec:res} as a reliable first-pass characterization of the corpus. Every response's full audit trail (each coder's codes and rationales, the adjudicator's ruling, and uncertainty flags) is preserved in the deposited materials, and a human validation study comparing an independently human-coded stratified sample
against the machine coding would be a valuable extension of this research effort. We note that using LLMs for coding data is an active and ongoing area of research in and of itself \citep{gilardi2023chatgpt,halterman2026codebook,ziems2024can}.

\bibliographystyle{unsrt}  
\bibliography{references}  

\end{document}

%% file: tables/codebook_item_19.tex
\begin{longtable}{p{1.5cm}p{2.6cm}p{3.4cm}p{7.5cm}}
\caption{AI Demo Feedback Post Qualitative Codebook}\label{tab:codebook_demo}\\
\toprule
Code ID & Category & Code & Description \\ \midrule
\endfirsthead
\toprule
Code ID & Category & Code & Description \\ \midrule
\endhead
\bottomrule
\endlastfoot
DF01 & Likes & Course Utility & Appreciation for demonstrations directly tied to class deliverables. Apply when the student praises the immediate, applied usefulness of demonstrated AI use for graded coursework (e.g., the truss project, homework, exams). \\
DF02 & Likes & Capability Awareness & Positive feedback regarding learning novel AI capabilities, prompting techniques, or new platforms (e.g., Gemini). Apply when the student expresses appreciation for discovering what AI can do or how to control it better, independent of a specific graded task. \\
DF03 & Likes & Broader Context & Enjoyment derived from seeing AI placed in larger real-world engineering, research, or industry context. Apply when the student praises the broader educational, societal, or research framing rather than immediate class tasks or tool mechanics. \\
DF04 & Likes & Other / Idiosyncratic Like & Valid, constructive positive feedback that does not fit DF01-DF03 and lacks sufficient frequency to form its own category. Describe the like in the notes field. \\
DF05 & Dislikes & Redundant / Basic & Frustration that the baseline AI literacy of the class was underestimated. Apply when the student expresses that they (or students generally) already knew the material presented. \\
DF06 & Dislikes & Lacked Course Utility & Frustration that the demos did not assist with immediate course hurdles (homework, core mechanics), including the feeling that demo time displaced core instruction. \\
DF07 & Dislikes & Logistics of Delivery Structure & Complaints about WHEN and HOW LONG: timing, placement in the lecture, duration, or scheduling of the demonstrations. \\
DF08 & Dislikes & Cognitive Overload \& Density & Frustration that the demonstrations were too complex, too dense, or induced mental fatigue: too much information to absorb, or content requiring background (e.g., programming) that students lacked. \\
DF09 & Dislikes & Low Engagement & Observations that the format failed to hold student attention: boredom, passive distraction, phone usage among peers, or self-reported disengagement. \\
DF10 & Dislikes & Other / Idiosyncratic Dislike & Valid, constructive negative feedback that does not fit DF05-DF09 and lacks sufficient frequency to form its own category. Describe the dislike in the notes field. \\
DF11 & Suggestions & Increase Course Utility & Recommendations to pivot demonstrations toward immediate academic help: exam preparation, generating practice problems, Socratic tutoring, project assistance, checking homework, or error-finding in student work. \\
DF12 & Suggestions & Increase Interactivity & Requests to move away from passive observation: guided, hands-on, class-wide prompting activities where students use the tools themselves, or more interactive formats generally. \\
DF13 & Suggestions & Modify Logistics of Delivery Structure & Suggestions to change how and when the demonstrations are delivered: moving demos to the start of class, making them shorter, or shifting them to optional asynchronous modules (e.g., Blackboard/Piazza video). \\
DF14 & Suggestions & Other / Idiosyncratic Suggestion & A valid, constructive suggestion that does not fit DF11-DF13 and lacks sufficient frequency to form its own category. Describe the suggestion in the notes field. \\
DF15 & Not Applicable & Did Not Attend / Cannot Comment & Student states they did not attend the demonstrations, attended too few to comment, or otherwise cannot evaluate them. Standalone when it is the whole response; if the student also comments substantively on demos they did see, code those comments and add DF15 only if partial attendance is explicitly the reason for limited feedback. \\
DF16 & Neutral & Neutral / No Strong Opinion & Student attended and reports no strong reaction in either direction, without substantive likes, dislikes, or suggestions. Standalone: if a specific like/dislike/suggestion is also given, code that instead. \\
DF17 & Uncodable & Irrelevant / Noise & Responses that are blank, nonsensical, or fail to address the prompt about the demonstrations in any capacity. Standalone. \\
\end{longtable}

%% file: tables/codebook_item_20.tex
\begin{longtable}{p{1.5cm}p{2.2cm}p{2.4cm}p{3.2cm}p{5.7cm}}
\caption{AI Workflow Steps Pre/Post Qualitative Codebook}\label{tab:codebook_workflow}\\
\toprule
Code ID & Tier & Category & Code & Description \\ \midrule
\endfirsthead
\toprule
Code ID & Tier & Category & Code & Description \\ \midrule
\endhead
\bottomrule
\endlastfoot
WF01 & Tier 1 & Process Structure & Self-Attempt First; Verify After & Student completes work independently, then uses AI to verify/check/validate the finished work. AI is post-hoc validation, not mid-problem help. \\
WF02 & Tier 1 & Process Structure & Self-Attempt First; AI Help When Stuck & Student attempts the problem independently FIRST and turns to AI only when stuck, confused, or unable to proceed; AI's role is mid-problem assistance (hints, explanations, next steps), after which the student resumes. \\
WF03 & Tier 1 & Process Structure & AI-Guided Step-by-Step & Student uses AI in a forward-facing, sequential manner from the outset: AI explains or generates steps and the student learns/works through them in linear progression. \\
WF04 & Tier 1 & Process Structure & Iterative Refinement & Multiple cycles of attempt-feedback-revision or dialogue-based refinement: the student returns to and revises previous work, or engages in multiple rounds of AI requests/prompt refinement to converge on understanding. \\
WF05 & Tier 1 & Process Structure & One-Shot Answer Generation & Student inputs the problem and takes the AI's answer/solution directly, with no independent attempt, no stepwise working-through, and no iteration described. \\
WF06 & Tier 1 & Process Structure & Non-Use / No Attempt & Explicit statement that the student does not use AI for this course, or does not attempt to. STANDALONE: select no Tier 2 codes (exception: WF18/WF19 may be applied when a limitation or distrust is stated as the reason for non-use). \\
WF22 & Tier 1 & Process Structure & Targeted Single-Purpose Use & Student names only a narrow, specific use of AI (e.g., concept reminders, debugging code, formula lookup) WITHOUT describing any sequential process from question to answer. The response describes what AI is for, not a workflow. Co-code the matching Tier 2 codes (often WF09, WF17). \\
WF07 & Tier 1 & Process Structure & Other Process Structure & A codable overall process structure is described but none of WF01-WF06 fits. Describe the structure in the notes field. \\
WF08 & Tier 1 & Uncodable & Irrelevant / Noise & Responses that provide no codable data: 'N/A', blank, nonsensical, or failing to address the prompt. STANDALONE: select no Tier 2 codes. \\
WF09 & Tier 2 & Conceptual Understanding \& Explanation & Concept Explanation Request & Student asks AI to explain concepts, theories, underlying principles, or abstract ideas: breaking down complex topics, simplifying, or clarifying examples. Focus on deepening comprehension of WHAT and WHY. \\
WF10 & Tier 2 & Procedural Learning \& Problem-Solving Steps & Procedure/Steps Request & Student asks AI for step-by-step procedures, equations, formulas, algorithms, methods, or sequential solutions: HOW to execute a process. \\
WF11 & Tier 2 & Procedural Learning \& Problem-Solving Steps & Procedural Memorization \& Practice & Student uses AI to memorize, internalize, drill, or practice procedures: flashcards, generated practice problems, quizzing, repetition toward fluency. \\
WF12 & Tier 2 & Procedural Learning \& Problem-Solving Steps & Guided Inquiry / Hints & Student asks AI for hints, guiding questions, leading prompts, or partial guidance rather than direct answers or full solutions; Socratic-style interaction intended to prompt thinking. \\
WF13 & Tier 2 & Verification, Checking, \& Error Correction & Answer/Result Verification & Student checks whether their FINAL answer or result is correct by comparing to AI's answer or having AI verify it. Quick 'is this right?' on the end product, not the process. \\
WF14 & Tier 2 & Verification, Checking, \& Error Correction & Comparative Step-by-Step Verification & Student solves independently, then reviews the AI's step-by-step solution to compare reasoning/methodology, examining HOW each step was done and where they match or differ. \\
WF15 & Tier 2 & Verification, Checking, \& Error Correction & Error Diagnosis / Troubleshooting & Student asks AI to identify and explain how and why errors occurred in the STUDENT'S work: root-cause analysis, debugging, locating where logic broke down. \\
WF16 & Tier 2 & Verification, Checking, \& Error Correction & Accuracy Verification Against External Sources & Student verifies correctness by comparing AI output against external authoritative sources: textbook, lecture notes, posted solutions, class problems. (Scope note: this codes verification steps mentioned within the workflow question; the verify\_correctness codebook applies only to the verification question.) \\
WF17 & Tier 2 & Strategic Approach \& Conditional Use & Conditional / Context-Dependent Use & Student explicitly describes conditions, boundaries, or contexts under which they use AI differently: only for certain topics, task types, or situations. \\
WF18 & Tier 2 & Tool Limitations \& Workarounds & Diagram / Visual Content Interpretation Issues & Student describes AI's inability or limitations in interpreting diagrams, schematics, images, or geometry, including workarounds (describing the figure in words, verifying the AI's reading of the figure). \\
WF19 & Tier 2 & Critical Evaluation \& Trust & Skepticism / Distrust of AI Outputs & Student expresses doubt, caution, or distrust about the reliability of AI outputs, including mentions of hallucination or frequent errors. \\
WF20 & Tier 2 & Prompting Strategy & Prompt Specification \& Constraint Setting & Student describes deliberately structuring or constraining prompts: withholding the answer request, instructing AI not to solve fully, specifying context/topic, isolating a sub-question, or refining inputs to control output. \\
WF21 & Tier 2 & Other & Other / Idiosyncratic Step & A specific, codable process step is described but no Tier 2 code fits. Describe it in the notes field. \\
\end{longtable}

%% file: tables/codebook_item_21.tex
\begin{longtable}{p{1.5cm}p{3.2cm}p{3.4cm}p{6.9cm}}
\caption{AI Verify Correctness Pre/Post Qualitative Codebook}\label{tab:codebook_verify}\\
\toprule
Code ID & Category & Code & Description \\ \midrule
\endfirsthead
\toprule
Code ID & Category & Code & Description \\ \midrule
\endhead
\bottomrule
\endlastfoot
A1 & Independent Verification & Re-Solves to Verify AI & Student independently works the problem (by hand, with a calculator, or step-by-step on their own) in order to verify the AI's response after receiving it. Solving the problem oneself ALONG or USING the AI's provided steps still counts as A1 when the student produces their own work; add A2 only if they also describe reading through the AI's reasoning as a separate check. \\
A2 & Independent Verification & Traces AI's Steps & Student reads through, follows, or walks through the AI's reasoning/derivation to look for errors, without independently re-deriving the solution. \\
A3 & Independent Verification & Plausibility / Sanity Check & Student describes an active plausibility judgment performed as a routine part of their workflow: common sense, physical intuition, units, magnitudes, directions. \\
A4 & Independent Verification & Verifies Diagram / Problem Setup & Student checks whether the AI correctly interpreted the problem inputs: the diagram or figure, geometry/positions, given values, or what is being asked. \\
A5 & Independent Verification & Solves First, Uses AI as Checker & Student's workflow is the reverse of A1: they complete the problem on their own FIRST, and the AI's role is to check the student's work. \\
B1 & External Cross-Reference & Course Materials \& Answer Keys & Student checks the AI response against course-provided resources: lecture notes/slides, the course textbook, posted solutions or answer keys. \\
B2 & External Cross-Reference & Instructor / TA Consultation & Student verifies the AI response (or the discrepancy between their answer and the AI's) by consulting course staff: professor, TA, LA, office hours. \\
B3 & External Cross-Reference & Peer Comparison & Student verifies by comparing answers or discussing the AI response with classmates, friends, or groupmates. \\
B4 & External Cross-Reference & Cross-Checks Other AI Models & Student runs the same question through one or more DIFFERENT AI tools/models (or re-runs the same tool for a fresh, independent response) and compares. \\
B5 & External Cross-Reference & Online Search & Student uses open-web resources that are NOT course-provided: Google searches, YouTube solution videos, online databases. \\
C1a & Within-Conversation AI Strategies & Direct Challenge / Error Flagging & Student asserts to the AI that something is (or may be) wrong: points out a suspected mistake, supplies a correction, or pushes back on a specific step. \\
C1b & Within-Conversation AI Strategies & Requests Re-Check / Redo & Student asks the AI to double-check, verify, recheck, or redo its own work, without asserting that a specific error exists. \\
C1c & Within-Conversation AI Strategies & Probing Follow-Ups & Student asks follow-up questions whose FUNCTION is to expose errors or test the AI's understanding, without directly asserting an error. Default code for generic, unspecified further prompting after a suspected issue ('if I find discontinuity, I prompt further') when the response does not state that an error was asserted (C1a) or a redo requested (C1b). \\
C2 & Within-Conversation AI Strategies & Verifies AI-Cited Sources & Student asks the AI to provide sources or citations and then independently checks those specific sources for accuracy. \\
D1 & Limited or No Verification & Minimal / Conditional Verification & Student describes checking that is conditional, rare, or passive: they check only when an answer seems off or surprising. \\
D2 & Limited or No Verification & No Verification & Student explicitly states they do not check AI responses at all, with no conditional or minimal checking described. \\
E1 & Not Applicable & Does Not Use AI & Student states they do not use AI tools, or have not used AI in this course, so the verification question does not apply. STANDALONE: never combine with any other code. \\
E2 & Not Applicable & Uncodable & Response is blank, 'N/A', 'yes/no' with no elaboration, 'don't know yet', nonsensical, or otherwise contains no identifiable verification behavior. STANDALONE: never combine with any other code. \\
\end{longtable}

%% file: tables/codebook_item_22.tex
\begin{longtable}{p{1.5cm}p{2.2cm}p{2.4cm}p{3.2cm}p{5.7cm}}
\caption{AI Impact Describe Pre/Post Qualitative Codebook}\label{tab:codebook_impact}\\
\toprule
Code ID & Tier & Category & Code & Description \\ \midrule
\endfirsthead
\toprule
Code ID & Tier & Category & Code & Description \\ \midrule
\endhead
\bottomrule
\endlastfoot
IM01 & Tier 1 & Overall Sentiment & Positive Impact & Student's net self-assessment is that the tool helped their learning. Apply when a net positive direction is stated ("positive", "helpful", "mostly positive"), even if drawbacks are also mentioned; capture drawbacks as Tier 2 drivers. \\
IM02 & Tier 1 & Overall Sentiment & Negative Impact & Student's net self-assessment is that the tool hindered their learning or was a net negative. \\
IM03 & Tier 1 & Overall Sentiment & Mixed Impact & Student weighs both pros and cons with active conflict or trade-offs and does NOT state a net direction, or explicitly says the impact is both positive and negative. \\
IM04 & Tier 1 & Overall Sentiment & Neutral / No Impact & Student USES (or has used) AI but reports it did not move the needle or had zero net effect. \\
IM05 & Tier 1 & Not Applicable & Does Not Use AI & Student states they do not use, or have not used, AI tools in this course, so the impact question does not apply to them. Speculation about future or hypothetical impact does not change this code. Tier 2 codes MAY be applied when the student gives concrete reasons for non-use that match a Tier 2 code (e.g., Technical Failure). \\
IM06 & Tier 1 & Uncodable & Irrelevant / Noise & Responses that provide no codable data: 'N/A', blank, nonsensical, or failing to address the prompt. Standalone; select no Tier 2 codes. \\
IM07 & Tier 2 & Positive Affordances & Concept Clarification & Use of AI to break down concepts, explain the 'bigger picture', or clarify confusing lecture topics. \\
IM08 & Tier 2 & Positive Affordances & Step Verification & Use of AI as a diagnostic tool to check intermediate steps, check answers, find mistakes, or confirm methodology. \\
IM09 & Tier 2 & Positive Affordances & On-Demand Accessibility & Valuation of AI for its efficiency, speed, or availability outside standard office hours. Includes dialogic availability: the ability to ask unlimited or immediate follow-up questions. \\
IM10 & Tier 2 & Negative Hindrances & Cognitive Reliance & Hindrance to the student's own cognitive process: reports of feeling 'lazy', over-reliant, losing critical thinking or problem-solving skill, or fear thereof from their own use. \\
IM11 & Tier 2 & Negative Hindrances & Technical Failure & Failure of the AI to perform the academic task correctly: wrong math, hallucination, inability to interpret mechanics diagrams or images. \\
IM12 & Tier 2 & Negative Hindrances & Pedagogical Misalignment & Frustration with the AI's teaching methods, explanation structure, or clarity, regardless of technical accuracy: strange strategies, misalignment with course material, vague or nonspecific explanations. \\
IM13 & Tier 2 & Conditional Use & Method-Dependent & Recognition that the tool's impact is tied to user behavior: AI is only positive 'if used correctly' and negative if misused. \\
IM15 & Tier 2 & Positive Affordances & Affective / Confidence Support & Emotional or confidence-related benefit: reduced frustration or stress, increased confidence, comfort asking questions without judgment, feeling supported or less stuck. \\
IM14 & Tier 2 & Emergent & Other / Idiosyncratic & A specific driver of sentiment is stated but fits no other Tier 2 code, even partially. Apply this code AND describe the novel driver in the notes field. A generic restatement of positivity or helpfulness is NOT a driver; check IM07-IM13 and IM15 first, and prefer a partial fit over this residual. \\
\end{longtable}

%% file: tables/codebook_item_23.tex
\begin{longtable}{p{1.5cm}p{2.6cm}p{3.4cm}p{7.5cm}}
\caption{AI Use Change Post Qualitative Codebook}\label{tab:codebook_usechange}\\
\toprule
Code ID & Category & Code & Description \\ \midrule
\endfirsthead
\toprule
Code ID & Category & Code & Description \\ \midrule
\endhead
\bottomrule
\endlastfoot
UC01 & No Substantive Change & No Substantive Change & AI use has not meaningfully changed over the semester, or the reported change is minimal/vague. Apply when the student reports continuity, or a slight shift without substantive detail. Standalone. \\
UC02 & Frequency of Use & Increased Frequency & Reported rise in how often AI is used, regardless of stated reason. \\
UC03 & Frequency of Use & Decreased Frequency & Reported decline in how often AI is used, including stopping entirely after prior use. CO-CODE with UC12 when the student names a non-AI replacement resource, and with UC10 when the stated reason is distrust or discovered limitations. \\
UC04 & Frequency of Use & Context-Dependent Change & Change in use that varies by subject, course, or problem difficulty rather than uniformly: more in some contexts, less in others. \\
UC05 & Manner of Use & Shift Toward Study \& Concept Support & Student's use has moved TOWARD learning-support tasks: studying, flashcards, concept review, or deepening comprehension. Direction rule: Manner of Use codes (UC05-UC07) describe the manner the student has adopted or increased, never a manner they report abandoning. \\
UC06 & Manner of Use & Shift Toward Solution \& Output Generation & Student's use has moved TOWARD output generation: writing code, solving problems, or generating text for deliverables. Same direction rule as UC05. \\
UC07 & Manner of Use & Shift Toward Verification \& Checking & Shift toward using AI to check or validate the student's own work rather than to generate it. Apply only when the student indicates doing the work themselves first. Same direction rule as UC05. \\
UC08 & Manner of Use & Prompting Skill Improvement & Improvement in how the student formulates queries or interacts with the tool: better prompting, providing more context, greater efficiency in getting useful responses. \\
UC09 & Stance \& Self-Regulation & Limiting Reliance & Deliberate moderation or restriction of AI to protect one's own learning: expressed concern about over-reliance, or intentionally limiting the task (e.g., withholding numbers) so the student still does the cognitive work. \\
UC10 & Stance \& Self-Regulation & Increased Critical Scrutiny & Growth in critical evaluation and awareness of AI limitations: questioning outputs, discovering poor performance in specific domains (e.g., trusses, diagrams), no longer trusting implicitly. \\
UC11 & Tool \& Resource Selection & Tool / Model Selection & Change in the AI tool, platform, or tier used: switching primary tools, using different models for different purposes, or changing subscription tiers. \\
UC12 & Tool \& Resource Selection & Shift to Non-AI Resources & Shift toward non-AI learning resources: instructors, office hours, TAs, textbooks, lecture notes, or videos in place of AI. Usually co-occurs with UC03. \\
UC13 & Other & Other / Unspecified Change & A valid, CLEAR change in AI use that fits none of the themes above and is too idiosyncratic to warrant its own code. Describe the change in the notes field. \\
UC14 & Uncodable & Irrelevant / Noise & Responses that are blank, nonsensical, or fail to address the prompt about changes in AI use in any capacity. Standalone. \\
UC15 & Not Applicable & Never Used AI & Student states they do not use AI at all and have not during the semester, so the change question does not apply. Standalone. \\
\end{longtable}

%% file: tables/table_reliability.tex
\begin{longtable}{lrrrrr}
\caption{Inter-model reliability of the LLM first-pass coding. Identical = percent of responses receiving identical code sets from all three models; $J$ = mean pairwise Jaccard similarity of code sets; $\tilde{\kappa}$ = median Fleiss' $\kappa$ across codes with mean prevalence $\geq$ 5\%; $\kappa \geq .61$ = percent of those codes reaching substantial agreement.}
\label{tab:llm_reliability} \\
\toprule
Question & $n$ & Identical (\%) & $J$ & $\tilde{\kappa}$ & $\kappa \geq .61$ (\%) \\
\midrule
Workflow steps (pre) & 99 & 52.5 & 0.81 & 0.82 & 92.3 \\
Workflow steps (post) & 99 & 38.4 & 0.76 & 0.79 & 93.8 \\
Verification behavior (pre) & 100 & 70.0 & 0.85 & 0.87 & 100.0 \\
Verification behavior (post) & 99 & 67.7 & 0.84 & 0.88 & 90.0 \\
Impact on learning (pre) & 100 & 70.0 & 0.90 & 0.91 & 92.3 \\
Impact on learning (post) & 99 & 70.7 & 0.91 & 0.90 & 90.9 \\
Demo feedback (post) & 60 & 71.7 & 0.89 & 0.90 & 100.0 \\
Change in AI use (post) & 99 & 81.8 & 0.92 & 0.92 & 100.0 \\
\bottomrule
\end{longtable}

%% file: tables/table_reliability_by_code.tex
{
\footnotesize
\begin{longtable}{llrrrr}
\caption{Per-code inter-model reliability. Prev.\ = mean prevalence across models; Final = prevalence in the adjudicated coding; $\kappa$ = Fleiss' $\kappa$ (shown when prevalence $\geq$ 5\%); Agr.\ = mean pairwise percent agreement.}
\label{tab:llm_reliability_by_code} \\
\toprule
Question & Code & Prev. & Final & $\kappa$ & Agr. \\
\midrule
Workflow steps (pre) & Self-Attempt First; Verify After & 0.20 & 0.21 & 0.86 & 0.95 \\
Workflow steps (pre) & Self-Attempt First; AI Help When Stuck & 0.26 & 0.24 & 0.86 & 0.95 \\
Workflow steps (pre) & AI-Guided Step-by-Step & 0.08 & 0.07 & 0.73 & 0.96 \\
Workflow steps (pre) & Iterative Refinement & 0.02 & 0.01 & -- & 0.99 \\
Workflow steps (pre) & One-Shot Answer Generation & 0.04 & 0.02 & -- & 0.97 \\
Workflow steps (pre) & Non-Use / No Attempt & 0.16 & 0.16 & 1.00 & 1.00 \\
Workflow steps (pre) & Targeted Single-Purpose Use & 0.11 & 0.14 & 0.67 & 0.93 \\
Workflow steps (pre) & Other Process Structure & 0.01 & 0.01 & -- & 0.97 \\
Workflow steps (pre) & Irrelevant / Noise & 0.11 & 0.14 & 0.97 & 0.99 \\
Workflow steps (pre) & Concept Explanation Request & 0.13 & 0.12 & 0.82 & 0.96 \\
Workflow steps (pre) & Procedure/Steps Request & 0.31 & 0.28 & 0.73 & 0.89 \\
Workflow steps (pre) & Procedural Memorization \& Practice & 0.05 & 0.04 & -- & 0.99 \\
Workflow steps (pre) & Guided Inquiry / Hints & 0.14 & 0.10 & 0.68 & 0.93 \\
Workflow steps (pre) & Answer/Result Verification & 0.21 & 0.21 & 0.84 & 0.95 \\
Workflow steps (pre) & Comparative Step-by-Step Verification & 0.10 & 0.10 & 0.81 & 0.97 \\
Workflow steps (pre) & Error Diagnosis / Troubleshooting & 0.10 & 0.10 & 0.92 & 0.99 \\
Workflow steps (pre) & Conditional / Context-Dependent Use & 0.03 & 0.02 & -- & 0.97 \\
Workflow steps (pre) & Diagram / Visual Content Interpretation Issues & 0.01 & 0.01 & -- & 0.99 \\
Workflow steps (pre) & Skepticism / Distrust of AI Outputs & 0.04 & 0.04 & -- & 0.99 \\
Workflow steps (pre) & Prompt Specification \& Constraint Setting & 0.07 & 0.08 & 0.59 & 0.95 \\
\midrule
Workflow steps (post) & Self-Attempt First; Verify After & 0.15 & 0.16 & 0.79 & 0.95 \\
Workflow steps (post) & Self-Attempt First; AI Help When Stuck & 0.20 & 0.20 & 0.79 & 0.93 \\
Workflow steps (post) & AI-Guided Step-by-Step & 0.20 & 0.20 & 0.79 & 0.93 \\
Workflow steps (post) & Iterative Refinement & 0.05 & 0.04 & -- & 0.97 \\
Workflow steps (post) & One-Shot Answer Generation & 0.04 & 0.04 & -- & 0.97 \\
Workflow steps (post) & Non-Use / No Attempt & 0.09 & 0.09 & 0.96 & 0.99 \\
Workflow steps (post) & Targeted Single-Purpose Use & 0.19 & 0.18 & 0.76 & 0.93 \\
Workflow steps (post) & Other Process Structure & 0.02 & 0.03 & -- & 0.97 \\
Workflow steps (post) & Irrelevant / Noise & 0.05 & 0.05 & 0.86 & 0.99 \\
Workflow steps (post) & Concept Explanation Request & 0.31 & 0.28 & 0.84 & 0.93 \\
Workflow steps (post) & Procedure/Steps Request & 0.37 & 0.36 & 0.73 & 0.87 \\
Workflow steps (post) & Procedural Memorization \& Practice & 0.02 & 0.02 & -- & 0.99 \\
Workflow steps (post) & Guided Inquiry / Hints & 0.08 & 0.06 & 0.64 & 0.95 \\
Workflow steps (post) & Answer/Result Verification & 0.19 & 0.19 & 0.89 & 0.97 \\
Workflow steps (post) & Comparative Step-by-Step Verification & 0.09 & 0.08 & 0.72 & 0.95 \\
Workflow steps (post) & Error Diagnosis / Troubleshooting & 0.06 & 0.05 & 0.66 & 0.96 \\
Workflow steps (post) & Accuracy Verification Against External Sources & 0.03 & 0.03 & -- & 0.99 \\
Workflow steps (post) & Conditional / Context-Dependent Use & 0.07 & 0.10 & 0.56 & 0.94 \\
Workflow steps (post) & Diagram / Visual Content Interpretation Issues & 0.09 & 0.08 & 0.92 & 0.99 \\
Workflow steps (post) & Skepticism / Distrust of AI Outputs & 0.11 & 0.11 & 0.83 & 0.97 \\
Workflow steps (post) & Prompt Specification \& Constraint Setting & 0.17 & 0.15 & 0.76 & 0.93 \\
Workflow steps (post) & Other / Idiosyncratic Step & 0.03 & 0.04 & -- & 0.99 \\
\midrule
Verification behavior (pre) & Re-Solves to Verify AI & 0.24 & 0.25 & 0.87 & 0.95 \\
Verification behavior (pre) & Traces AI's Steps & 0.30 & 0.34 & 0.83 & 0.93 \\
Verification behavior (pre) & Plausibility / Sanity Check & 0.17 & 0.13 & 0.70 & 0.91 \\
Verification behavior (pre) & Verifies Diagram / Problem Setup & 0.03 & 0.03 & -- & 1.00 \\
Verification behavior (pre) & Solves First, Uses AI as Checker & 0.06 & 0.06 & 0.76 & 0.97 \\
Verification behavior (pre) & Course Materials \& Answer Keys & 0.14 & 0.13 & 0.92 & 0.98 \\
Verification behavior (pre) & Instructor / TA Consultation & 0.01 & 0.01 & -- & 1.00 \\
Verification behavior (pre) & Peer Comparison & 0.11 & 0.11 & 1.00 & 1.00 \\
Verification behavior (pre) & Cross-Checks Other AI Models & 0.10 & 0.10 & 1.00 & 1.00 \\
Verification behavior (pre) & Online Search & 0.07 & 0.08 & 0.90 & 0.99 \\
Verification behavior (pre) & Direct Challenge / Error Flagging & 0.02 & 0.02 & -- & 1.00 \\
Verification behavior (pre) & Requests Re-Check / Redo & 0.00 & 0.00 & -- & 0.99 \\
Verification behavior (pre) & Probing Follow-Ups & 0.03 & 0.02 & -- & 1.00 \\
Verification behavior (pre) & Verifies AI-Cited Sources & 0.02 & 0.02 & -- & 1.00 \\
Verification behavior (pre) & Minimal / Conditional Verification & 0.04 & 0.05 & -- & 0.97 \\
Verification behavior (pre) & No Verification & 0.02 & 0.02 & -- & 0.99 \\
Verification behavior (pre) & Does Not Use AI & 0.03 & 0.03 & -- & 1.00 \\
Verification behavior (pre) & Uncodable & 0.11 & 0.09 & 0.82 & 0.97 \\
\midrule
Verification behavior (post) & Re-Solves to Verify AI & 0.29 & 0.30 & 0.89 & 0.95 \\
Verification behavior (post) & Traces AI's Steps & 0.23 & 0.26 & 0.76 & 0.91 \\
Verification behavior (post) & Plausibility / Sanity Check & 0.08 & 0.05 & 0.50 & 0.93 \\
Verification behavior (post) & Verifies Diagram / Problem Setup & 0.08 & 0.07 & 0.86 & 0.98 \\
Verification behavior (post) & Solves First, Uses AI as Checker & 0.05 & 0.04 & -- & 0.99 \\
Verification behavior (post) & Course Materials \& Answer Keys & 0.17 & 0.16 & 0.98 & 0.99 \\
Verification behavior (post) & Instructor / TA Consultation & 0.01 & 0.01 & -- & 1.00 \\
Verification behavior (post) & Peer Comparison & 0.08 & 0.08 & 1.00 & 1.00 \\
Verification behavior (post) & Cross-Checks Other AI Models & 0.09 & 0.09 & 0.96 & 0.99 \\
Verification behavior (post) & Online Search & 0.14 & 0.14 & 0.97 & 0.99 \\
Verification behavior (post) & Direct Challenge / Error Flagging & 0.03 & 0.02 & -- & 0.99 \\
Verification behavior (post) & Requests Re-Check / Redo & 0.03 & 0.03 & -- & 0.99 \\
Verification behavior (post) & Probing Follow-Ups & 0.05 & 0.06 & -- & 0.98 \\
Verification behavior (post) & Verifies AI-Cited Sources & 0.01 & 0.01 & -- & 0.99 \\
Verification behavior (post) & Minimal / Conditional Verification & 0.12 & 0.12 & 0.72 & 0.94 \\
Verification behavior (post) & No Verification & 0.01 & 0.01 & -- & 0.99 \\
Verification behavior (post) & Does Not Use AI & 0.03 & 0.03 & -- & 1.00 \\
Verification behavior (post) & Uncodable & 0.06 & 0.05 & 0.88 & 0.99 \\
\midrule
Impact on learning (pre) & Positive Impact & 0.47 & 0.47 & 0.93 & 0.97 \\
Impact on learning (pre) & Negative Impact & 0.06 & 0.07 & 0.88 & 0.99 \\
Impact on learning (pre) & Mixed Impact & 0.10 & 0.11 & 0.82 & 0.97 \\
Impact on learning (pre) & Neutral / No Impact & 0.10 & 0.10 & 0.89 & 0.98 \\
Impact on learning (pre) & Does Not Use AI & 0.17 & 0.15 & 0.91 & 0.97 \\
Impact on learning (pre) & Irrelevant / Noise & 0.09 & 0.10 & 0.96 & 0.99 \\
Impact on learning (pre) & Concept Clarification & 0.31 & 0.30 & 0.89 & 0.95 \\
Impact on learning (pre) & Step Verification & 0.05 & 0.05 & 0.86 & 0.99 \\
Impact on learning (pre) & On-Demand Accessibility & 0.09 & 0.09 & 0.96 & 0.99 \\
Impact on learning (pre) & Cognitive Reliance & 0.15 & 0.14 & 0.92 & 0.98 \\
Impact on learning (pre) & Technical Failure & 0.05 & 0.05 & 1.00 & 1.00 \\
Impact on learning (pre) & Pedagogical Misalignment & 0.02 & 0.02 & -- & 0.99 \\
Impact on learning (pre) & Method-Dependent & 0.09 & 0.09 & 0.92 & 0.99 \\
Impact on learning (pre) & Affective / Confidence Support & 0.01 & 0.00 & -- & 0.99 \\
Impact on learning (pre) & Other / Idiosyncratic & 0.06 & 0.09 & 0.55 & 0.95 \\
\midrule
Impact on learning (post) & Positive Impact & 0.60 & 0.59 & 0.89 & 0.95 \\
Impact on learning (post) & Negative Impact & 0.05 & 0.04 & 1.00 & 1.00 \\
Impact on learning (post) & Mixed Impact & 0.14 & 0.16 & 0.76 & 0.94 \\
Impact on learning (post) & Neutral / No Impact & 0.16 & 0.17 & 0.95 & 0.99 \\
Impact on learning (post) & Does Not Use AI & 0.03 & 0.03 & -- & 1.00 \\
Impact on learning (post) & Irrelevant / Noise & 0.02 & 0.01 & -- & 0.99 \\
Impact on learning (post) & Concept Clarification & 0.43 & 0.43 & 0.94 & 0.97 \\
Impact on learning (post) & Step Verification & 0.05 & 0.05 & 0.93 & 0.99 \\
Impact on learning (post) & On-Demand Accessibility & 0.17 & 0.19 & 0.85 & 0.96 \\
Impact on learning (post) & Cognitive Reliance & 0.11 & 0.10 & 0.90 & 0.98 \\
Impact on learning (post) & Technical Failure & 0.09 & 0.09 & 1.00 & 1.00 \\
Impact on learning (post) & Pedagogical Misalignment & 0.06 & 0.06 & 0.78 & 0.97 \\
Impact on learning (post) & Method-Dependent & 0.05 & 0.04 & 0.60 & 0.96 \\
Impact on learning (post) & Affective / Confidence Support & 0.01 & 0.01 & -- & 1.00 \\
Impact on learning (post) & Other / Idiosyncratic & 0.03 & 0.03 & -- & 0.97 \\
\midrule
Demo feedback (post) & Course Utility & 0.16 & 0.17 & 0.96 & 0.99 \\
Demo feedback (post) & Capability Awareness & 0.13 & 0.13 & 0.90 & 0.98 \\
Demo feedback (post) & Broader Context & 0.05 & 0.05 & 1.00 & 1.00 \\
Demo feedback (post) & Other / Idiosyncratic Like & 0.07 & 0.07 & 0.75 & 0.97 \\
Demo feedback (post) & Redundant / Basic & 0.12 & 0.12 & 1.00 & 1.00 \\
Demo feedback (post) & Lacked Course Utility & 0.18 & 0.18 & 0.89 & 0.97 \\
Demo feedback (post) & Logistics of Delivery Structure & 0.06 & 0.03 & 0.71 & 0.97 \\
Demo feedback (post) & Cognitive Overload \& Density & 0.07 & 0.07 & 1.00 & 1.00 \\
Demo feedback (post) & Low Engagement & 0.09 & 0.08 & 0.79 & 0.97 \\
Demo feedback (post) & Other / Idiosyncratic Dislike & 0.07 & 0.07 & 0.82 & 0.98 \\
Demo feedback (post) & Increase Course Utility & 0.22 & 0.22 & 0.90 & 0.97 \\
Demo feedback (post) & Increase Interactivity & 0.09 & 0.10 & 0.94 & 0.99 \\
Demo feedback (post) & Modify Logistics of Delivery Structure & 0.12 & 0.12 & 0.95 & 0.99 \\
Demo feedback (post) & Other / Idiosyncratic Suggestion & 0.13 & 0.13 & 0.85 & 0.97 \\
Demo feedback (post) & Neutral / No Strong Opinion & 0.02 & 0.02 & -- & 1.00 \\
Demo feedback (post) & Irrelevant / Noise & 0.06 & 0.07 & 0.89 & 0.99 \\
\midrule
Change in AI use (post) & No Substantive Change & 0.39 & 0.41 & 0.93 & 0.97 \\
Change in AI use (post) & Increased Frequency & 0.08 & 0.07 & 0.91 & 0.99 \\
Change in AI use (post) & Decreased Frequency & 0.09 & 0.10 & 0.96 & 0.99 \\
Change in AI use (post) & Context-Dependent Change & 0.03 & 0.02 & -- & 0.98 \\
Change in AI use (post) & Shift Toward Study \& Concept Support & 0.11 & 0.11 & 0.90 & 0.98 \\
Change in AI use (post) & Shift Toward Solution \& Output Generation & 0.04 & 0.03 & -- & 0.98 \\
Change in AI use (post) & Shift Toward Verification \& Checking & 0.05 & 0.04 & -- & 0.99 \\
Change in AI use (post) & Prompting Skill Improvement & 0.04 & 0.04 & -- & 0.99 \\
Change in AI use (post) & Limiting Reliance & 0.06 & 0.04 & 0.75 & 0.97 \\
Change in AI use (post) & Increased Critical Scrutiny & 0.10 & 0.10 & 0.89 & 0.98 \\
Change in AI use (post) & Tool / Model Selection & 0.08 & 0.08 & 1.00 & 1.00 \\
Change in AI use (post) & Shift to Non-AI Resources & 0.03 & 0.03 & -- & 1.00 \\
Change in AI use (post) & Other / Unspecified Change & 0.03 & 0.02 & -- & 0.98 \\
Change in AI use (post) & Irrelevant / Noise & 0.06 & 0.06 & 1.00 & 1.00 \\
Change in AI use (post) & Never Used AI & 0.02 & 0.02 & -- & 1.00 \\
\bottomrule
\end{longtable}
}